\documentclass[11pt,amsmath,amssymb,aps,nofootinbib]{revtex4-1}
\usepackage{graphicx,epsfig,dcolumn,bm,epic,eepic,float}
\usepackage{amsmath}
\usepackage{latexsym}
\usepackage{color}
\usepackage{cancel}
\usepackage{makeidx,shortvrb,latexsym}
\usepackage{epstopdf}
\usepackage{graphicx}

\pdfoutput=1

\begin{document}
\noindent{\hfill \small IPPP/19/78}

\noindent{\hfill \small MAN/HEP/2019/006
\\[0.1in]
}
\title{$k_t$-factorization vs collinear factorization in $W^{+}W^{-}$ pair production at the LHC}

\author{N. Darvishi\footnote{neda.darvishi@manchester.ac.uk}, 
        M.R. Masouminia\footnote{mohammad.r.masouminia@durham.ac.uk} and 
        K. Ostrolenk\footnote{kiran.ostrolenk@postgrad.manchester.ac.uk}} 
\affiliation{\it \small
$^{a,c}$School of Physics and Astronomy, University of Manchester, Oxford Road, Manchester M13 9PL, United Kingdom
\\
$^{b}$Institute for Particle Physics Phenomenology, Department of Physics, Durham University, South Road, Durham DH1 3LE, United Kingdom
}

\begin{abstract}

In this paper, we calculate the inclusive rate of $W^{+} W^{-}$ pair production through leptonic decay channels $W^+W^- \to l^+\nu_l + l^{\prime -} \nu_{l'}$ in the $k_t$-factorization framework. We also consider the exclusive $W^{+} W^{-}$ pair production through the one-loop induced $gg \to H \to W^+W^-$ channel that is important for the study of new physics beyond the Standard Model. The results are compared with predictions from the \textsf{Herwig~7} event generator in the collinear factorization framework and with the experimental data from the ATLAS and the CMS collaborations. It will be shown that our predictions for the $W^+W^-$ boson pair production signals are in agreement with the experimental data as well as the collinear results.

\end{abstract}

\maketitle

\section{Introduction}
\label{sec:intro}

The production of the $W^{+} W^{-}$ pairs is one of the most important electroweak (EW) processes at the CERN Large Hadron Collider (LHC), that provides a strong test for the viability of the Standard Model (SM) \cite{Abazov:2009ys, Abazov:2009tr, Aaltonen:2009aa, Chatrchyan:2011tz, Aad:2011kk, Abazov:2011cb, Aad:2012rxl, ATLAS:2012mec, Chatrchyan:2013oev, Chatrchyan:2013yaa, ATLAS:2014xea, Khachatryan:2015sga, Aad:2016wpd, Aaboud:2017qkn, Gallo:2018agq, Cuevas:2018jah, Aaboud:2019nkz, Aaboud:2018jqu, Aad:2019lpq, Aad:2016lvc, Khachatryan:2016vnn, Aaboud:2019nkz}. Moreover, these processes have played a key role in precision measurements at the LHC and also for the estimation of the irreducible backgrounds in Higgs boson searches. Furthermore, detecting any deviation from the theoretical predictions of the SM could automatically signal the existence of new physics beyond the Standard Model\,(BSM). This is even more relevant for heavy gauge vector boson (HGVB) pair production events (e.g.~$\gamma / H/Z^0 \to W^+ W^-$), due to their sensitivity to BSM modifications, e.g. via an extended Higgs sector~\cite{Darvishi:2016fwo}.

To the date, a number of theoretical calculations for the prediction of the $W^+W^-$ pair production rate have been done in different frameworks, to LO \cite{Meade:2014fca}, NLO \cite{Bai:2012zza,Wang:2013qua,Campanario:2013wta,Bellm:2016cks} and NNLO \cite{Gehrmann:2014fva,Grazzini:2016ctr} in QCD accuracies. 

Here, we calculate the inclusive rate of $W^{+} W^{-}$ pair production through leptonic decay channels $W^+W^- \to l^+\nu_l + l^{\prime -} \nu_{l'}$ in two different approaches: 
\\
(i) \textbf{The $\bf k_t$-factorization framework} where we use the transverse momentum dependent parton distribution functions (TMD PDFs) of Kimber-Martin-Ryskin (KMR) \cite{Kimber:2001sc,Martin:2009ii}.
In this framework, un-integrated parton distributions (UPDFs) are used to weight the relevant partonic sub-processes. These transverse-momentum-dependent parton distribution functions (TMD PDFs) are defined based on the solutions of the Dokshitzer-Gribov-Lipatov-Altarelli-Parisi (DGLAP) evolution equations \cite{Gribov:1972ri,Lipatov:1974qm,Altarelli:1977zs,Dokshitzer:1977sg}, and the last-step evolution approximation. The latter, directly introduces transverse momentum dependency into the partonic densities by softening the strong ordering assumption, i.e. $k^2_{t,1} \ll \cdots \ll k^2_{t,\text{semi-hard}} \sim \mu^2_{\text{hard}}$. This formalism suppresses the soft gluon singularities that arise from color coherence interference effects, by employing the angular ordering constraint (AOC) \cite{Deak:2015dpa} and effectively limits its evolution to a singularity-free phase space. Hence, the choice of AOC has a definitive effect on the characteristics of the resulting UPDF. Employing different AOC visualizations have resulted in different formalisms, namely the KMR, LO Martin-Ryskin-Watt (MRW) and NLO MRW UPDFs \cite{Martin:2009ii}. The capability of these phenomenological assets in correctly describing various physical observables have been already investigated, see e.g. \cite{Darvishi:2016fwo,Modarres:2016tow,Modarres:2016hpe,Modarres:2016phz,Modarres:2018dwj,AminzadehNik:2018kch}. 
\\
(ii) \textbf{The collinear factorization framework} which can be generated using the \textsf{Herwig 7} (v7.1.5) event generator \cite{Bahr:2008pv,Bellm:2015jjp,Bellm:2017bvx}. The corresponding matrix elements are evaluated by \textsf{MadGraph5} \cite{Alwall:2014hca} and \textsf{OpenLoops} packages \cite{Cascioli:2011va,Buccioni:2017yxi,Kallweit:2014xda} for the tree-level and QCD one-loop-induced channels, respectively. 
NLO QCD corrections have been calculated using \textsf{Matchbox} \cite{Platzer:2011bc,Bellm:2019wrh,Bellm:2020} and the partonic sub-processes are showered by a QCD+QED parton shower which is \textit{angularly ordered} and \textit{MC@NLO matched} \cite{Frixione:2002ik}. Finally, hadronization was performed using the cluster model of \cite{Webber:1983if} and the results have been analyzed using \textsf{Rivet} \cite{Buckley:2010ar}\footnote{The \textsf{Herwig 7} multi-purpose event generator provides a versatile platform for performing this class of analysis. Such calculations can be made in different setups, depending on the requirements, for instance, see~\cite{Bellm:2016cks}.}.

 Additionally, we calculate the exclusive $W^{+} W^{-}$ pair production through the one-loop induced $gg \to H \to W^+W^-$ channel in both $k_t$-factorization and collinear frameworks. This channel is particularly important in searching for BSM signal at the LHC.
We will report our study on exclusive $W^{+} W^{-}$ pair production in BSM within an upcoming paper. There we will use the BSM sensitive $gg \to H_i \to W^+ W^-$ vertices, with $H_i$ being either the SM Higgs boson or extra Higgs scalars in the Maximally Symmetric Two Higgs Doublet Model (MS-2HDM) \cite{Darvishi:2019ltl} and in the 2HDM type II with cancellation quadratic divergences \cite{Darvishi:2017bhf}. We will compare our results for both inclusive and exclusive $W^{+} W^{-}$ pair production with the existing experimental data from  ATLAS \cite{Aad:2016wpd, Aad:2016lvc,Aaboud:2019nkz} and CMS \cite{Khachatryan:2016vnn}.

The outline of this paper is as follows: Section \ref{sec:ww} includes the details of partonic interactions for $W^{+} W^{-}$ pair production at the LHC. In Section \ref{sec:Framework}, we present a brief description of our calculations in the $k_t$-factorization framework and will show the setups for our numerical analysis. Section \ref{sec:Results} contains our predictions for the inclusive $W^+ W^-$ pair production as well as the Higgs-decay-driven $W^+ W^-$ production. The results of both frameworks are compared with the experimental data from the ATLAS and the CMS collaborations. These comparisons will show that our predictions for the $W^+W^-$ boson pair production signals are in agreement with the experimental data. Finally, the section \ref{sec:Conclusions} will outline our conclusions.

\section{$W^{+} W^{-}$ pair production}
\label{sec:ww}

\begin{figure}
\centering
\includegraphics[width=.8\textwidth]{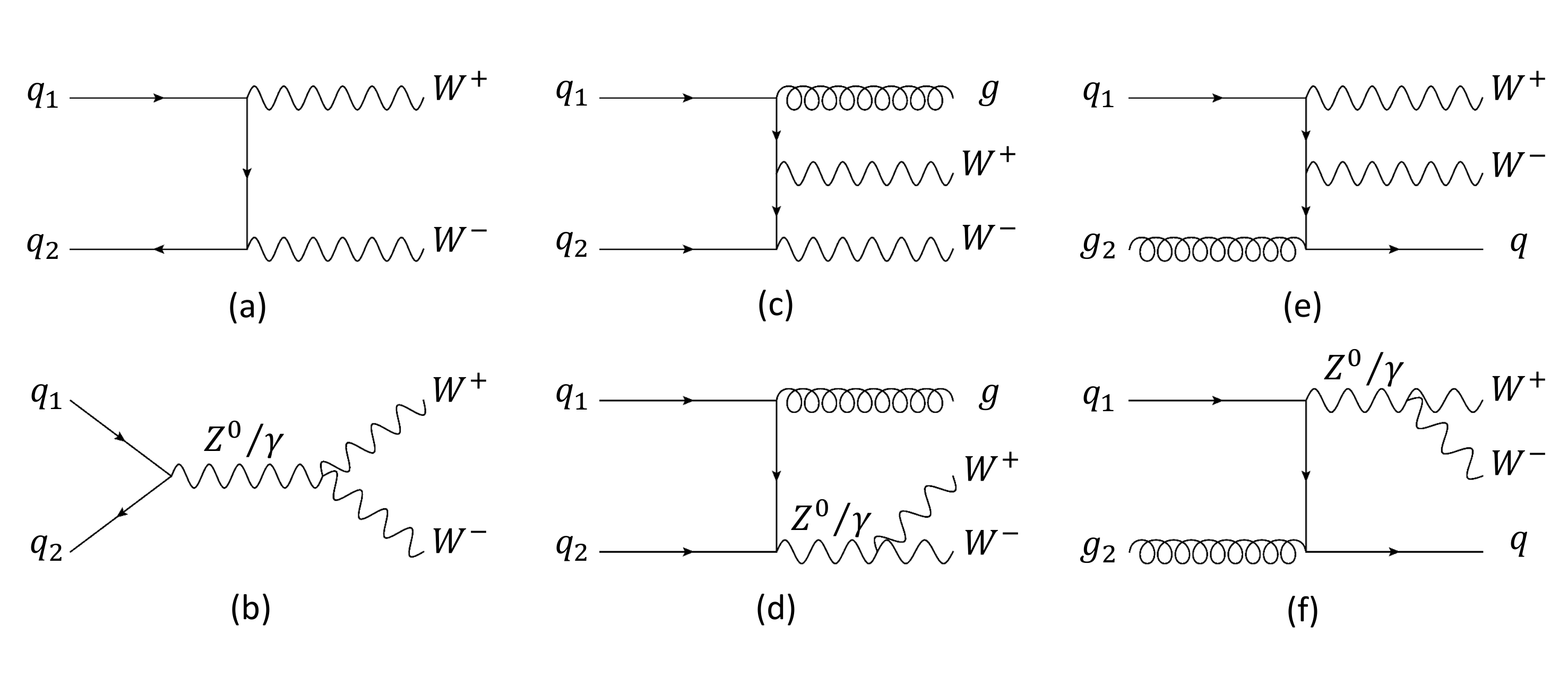}
\caption{ \it Tree-level contributions (up to one jet) in the cross-section for the production of $W^{+} W^{-}$ pairs~at~the~LHC.}
\label{fig1}
\end{figure}

\begin{figure}
\centering
\includegraphics[width=0.8\textwidth]{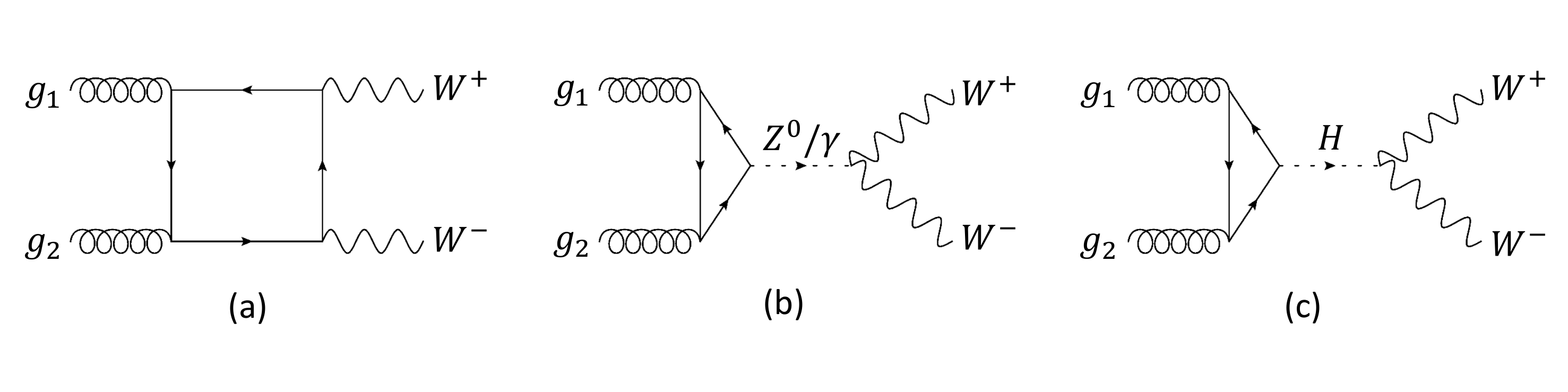}
\caption{ \it One-loop induced $g+g \to W^+ + W^-$ channels in $W^{+} W^{-}$ pair production~at~the~LHC.}
\label{fig2}
\end{figure}

We consider the inclusive production of the $W^{+} W^{-}$ pairs at the LHC that decay to two charged leptons and two neutrinos. In such processes, the general hadronic scattering is
\begin{eqnarray}
A(P_1) + B(P_2) & \to & a(k_1) + b(k_2) \to W^{+}(q_1) + W^{-}(q_2) + X
\nonumber \\
& \to & l^{+}(p_1) + \nu_l(p_2) + l^{-}(p_3)+ \bar{\nu}_l(p_4)+ X,
\label{PP_WW}
\end{eqnarray}
with $A(P_1)$ and $B(P_2)$ being the colliding beam protons, $a(k_1)$ and $b(k_2)$ denote the partons that initiate any of the involved sub-processes and $ l^{\pm}=e^{\pm},\mu^{\pm}$ are the resulting final-state fermions. The 4-momenta of the beam protons, the initial partons, the exchanged $W$-bosons and the final state leptons are denoted by $P_i$, $k_j$, $q_m$ and $p_n$, respectively. The hadronic scattering (\ref{PP_WW}) is dominated by the following partonic channels:
\begin{eqnarray}
q_1+q_2 \to & W^{+}+W^{-} &, \label{J2WW} \\
q_1+q_2 \to Z^0/\gamma \to & W^{+}+W^{-} &, \label{Z2WW} \\ 
q_1+q_2 \to & W^{+}+W^{-} & +J, \label{qq2WWj} \\
q_1+q_2 \to Z^0/\gamma +J \to & W^{+}+W^{-} & +J, \label{qq2Z2WWj} \\
q_1+g_2 \to & W^{+}+W^{-} & +J, \label{qg2WWj} \\
q_1+g_2 \to Z^0/\gamma +J \to & W^{+}+W^{-} & +J, \label{qg2Z2WWj} \\
g_1+g_2 \to & W^{+}+W^{-} &, \label{gg-loop} \\
g_1+g_2 \to Z^0/\gamma \to & W^{+}+W^{-} &, \label{gg-Z-loop} \\
g_1 + g_2 \to H \to & W^{+}+W^{-} &. \label{H2WW}
\end{eqnarray}
These sub-processes are depicted in Figure~\ref{fig1} and Figure~\ref{fig2}. 

The relevant leading-order (LO) sub-processes are comprised only of the quark-quark tree-level contributions (\ref{J2WW}) and the LO Drell-Yan type processes (\ref{Z2WW}), which are the dominant contributions into the inclusive $W^+ W^-$ pair production events at the LHC current center-of-mass energies. These sub-processes are shown as the diagrams (a) and (b) of the Figure~\ref{fig1}. Additionally, the relevant tree-level plus one jet sub-processes are the $P_1+P_2~\to~W^{+}+W^{-}+J$ contributions shown in processes (\ref{qq2WWj}), (\ref{qq2Z2WWj}),(\ref{qg2WWj}) and (\ref{qg2Z2WWj}) and correspondingly in the diagrams (c) through (f) of the Figure~\ref{fig1}. 

In the case of the $W^+W^-$ pair production via QCD one-loop channels, one can argue the importance of counting the gluon-gluon fusion contributions via single QCD loop diagrams as shown in sub-processes (\ref{gg-loop}) and (\ref{gg-Z-loop}) and correspondingly in the diagrams (a) and (b) of the Figure~\ref{fig2} \cite{Grazzini:2016ctr,Caola:2015rqy,Gehrmann:2014fva}. This, despite being valid in the case of the collinear framework, will cause irreducible double-counting while working within the $k_t$-factorization formalism, directly due to the use of the box and crossed-box diagrams in the definition of the $k_t$-factorization partonic densities\footnote{\tiny This argument is also valid in the case of one-loop quark-quark fusion channels via a single QCD loop. If one is interested in counting these types of contributions in the current production events, one has to omit the LO channels and only sum over the full single-loop NLO range, including the gluon-gluon and quark-quark fusion channels that have the same number of jets in their final states \cite{Modarres:2018dwj}. Hence, the set of sub-processes (\ref{J2WW}) through (\ref{qg2Z2WWj}) plus the sub-process (\ref{H2WW}) are the QCD NLO-level-equivalent sub-processes for the inclusive calculation of $W^+W^-$ pair in the $k_t$-factorization framework.} \cite{Watt:2003vf,Watt:2003mx, Modarres:2016tow}.

The production of $W^+ W^-$ pairs through Higgs decay is predominantly dominated by the gluon-gluon fusion channel as shown in the sub-process (\ref{H2WW}) and in Figure~\ref{fig2}c. Note that the higher-order (QCD and radiative) corrections to this channel are considerable. However, it has been shown that by using the K-factor approximation one can capture a very good description of the event, without the need for adding higher-order calculations \cite{Campbell:2006wx, Modarres:2018dwj}. 

The existence of a leptonic final-state for a $W^+ W^-$ pair production event provides a clean experimental signature while preventing reconstruction of the $H \to W^+ W^-$ resonance. The latter, in turn, causes some difficulty for the detection of the background signals and lowers the sensitivity of the experimental measurements \cite{Grazzini:2016ctr}. Therefore, it would be more important to increase the precision of the theoretical calculations to enhance our understanding of the $H \to W^+ W^-$ signal \cite{Campbell:2013wga,Frixione:1993yp}. Meanwhile, since it is not possible to calculate the invariant masses of the HGVB in a pair-production event, the process (\ref{PP_WW}) can be considered as an irreducible background signal for many LHC measurements. 

\section{Calculation Framework}
\label{sec:Framework}

At the hadronic level, the differential cross-section for the production of $W^+ W^-$ pairs (via leptonic decay) in the $k_t$-factorization framework can be written as
\begin{widetext}
\begin{eqnarray}
    d\sigma(AB \to l^{+} \nu_l  l^{-} \bar{\nu}_l+[j]) &=& 
    \sum_{a,b=q,g} \int {dx_1 \over x_1} {dx_2 \over x_2}
    {dk_{1,t}^2 \over k_{1,t}^2} {dk_{2,t}^2 \over k_{2,t}^2}
    \; f_{a}(x_1,k_{1,t}^2,\mu^2)\; f_{b}(x_2,k_{2,t}^2,\mu^2) \; 
    \nonumber \\ &\times&
    {d\phi(ab \to l^{+} \nu_l  l^{-} \bar{\nu}_l+[j]) \over F_{AB \to ab}}
	|{\mathcal{M}}(ab \to W^+W^- \to l^{+} \nu_l  l^{-} \bar{\nu}_l+[j])|^2 ,
	\nonumber \\
    \label{TCS}
\end{eqnarray}
\end{widetext}
with the particle phase space, $d\phi$, and the flux factor, $F$, defined as
	\begin{eqnarray}	
	d\phi &=& (2\pi)^4
	\prod_{i\in \text{final-state}} \left[ {1 \over 16\pi^2} \; dp_{i,t}^2 \; dy_i \; {d\phi_i \over 2\pi} \right]
	\delta^{(4)} \left( k_1 + k_2 -\sum_{j\in \text{final-state}} p_j \right) , 
    \label{dPHI}
	\end{eqnarray}
	\begin{equation}	
	F = x_1 x_2 s.
    \label{flux}
	\end{equation}
Here, $y_i$ and $\phi_i$ are the pseudorapidities and the azimuthal angles of emission of the final-state particles, respectively, while $s$ in the Eq. (\ref{flux}) is the center-of-mass energy squared in the infinite momentum frame, $P_i^2 \gg m^2_{proton}$,
\begin{eqnarray}
s &=& (P_1 + P_2)^2 = 2P_1 \cdot P_2.
\end{eqnarray}
Also, $x_{1}$ and $x_{2}$ are defined as
\begin{eqnarray}
x_{1,2} = {1 \over \sqrt{s}} \sum_{i\in \text{final-state}}  m_{i,t} e^{\pm y_i},
\end{eqnarray}
with $m_{i,t} = (m_i^2 + p_{i,t}^2)^{1/2}$ being the transverse mass of the final state particles. 
The matrix elements of the relevant sub-processes, $\mathcal{M}$, are calculated using the Feynman rules in combination with the \textit{Eikonal approximation} for the incoming quark spin densities and the \textit{non-sense polarization approximation} for the polarization vectors of the incoming gluons \cite{Modarres:2018dwj,Deak:thesis,Baranov:2008hj,Levin:1991ry,Gribov:1984tu,Catani:1990eg,Collins:1991ty}. To generate the analytic expressions for these matrix elements, we have used the algebraic manipulation package \textsf{FORM} \cite{FORM} and checked our results, independently, by \textsf{MATHEMATICA}. 

The differential cross-section for $W^+ W^-$ pair production of hadronic collisions can be derived for the sub-processes (\ref{J2WW}) and (\ref{Z2WW}) as  

\begin{widetext}
\begin{eqnarray}
d\sigma(AB \to l^{+} \nu_l  l^{-} \bar{\nu}_l) &=& \sum_{a,b=q,g} {dk_{1,t}^2 \over k_{t,1}^2} {dk_{2,t}^2 \over k_{2,t}^2} \; dp_{1,t}^2 \; dp_{2,t}^2 \; dp_{3,t}^2 
 \; dy_1 \; dy_2 \; dy_3 \; dy_4 \; {d\psi_1 \over 2\pi} {d\psi_2 \over 2\pi} {d\phi_1 \over 2\pi} {d\phi_2 \over 2\pi} {d\phi_3 \over 2\pi}
\nonumber \\ & \times & 
f_{a}(x_1,k_{1,t}^2, \mu^2) f_{b}(x_2,k_{2,t}^2, \mu^2) {|\mathcal{M}(ab \to W^+W^- \to l^{+} \nu_l  l^{-} \bar{\nu}_l)|^2 \over 2024\pi^4 (x_1 x_2 s)^2},
\label{TCS_224}
\end{eqnarray}
and for the sub-processes (\ref{qq2WWj}) through (\ref{qg2Z2WWj}) as
\begin{eqnarray}
d\sigma(AB \to l^{+} \nu_l  l^{-} \bar{\nu}_l+j) &=& \sum_{a,b=q,g} {dk_{1,t}^2 \over k_{1,t}^2} {dk_{2,t}^2 \over k_{2,t}^2} \; dp_{1,t}^2 \; dp_{2,t}^2 \; dp_{3,t}^2 \; dp_{4,t}^2
 \; dy_1 \; dy_2 \; dy_3 \; dy_4 \; dy_5 \; 
 \nonumber \\ & \times & 
 {d\psi_1 \over 2\pi} {d\psi_2 \over 2\pi} {d\phi_1 \over 2\pi} 
 {d\phi_2 \over 2\pi} {d\phi_3 \over 2\pi} {d\phi_4 \over 2\pi}
 f_{a}(x_1,k_{1,t}^2, \mu^2) f_{b}(x_2,k_{2,t}^2, \mu^2)
\nonumber \\ & \times & 
{|\mathcal{M}(ab \to W^+W^- \to l^{+} \nu_l  l^{-} \bar{\nu}_l+j)|^2 \over 32768\pi^5 (x_1 x_2 s)^2}.
\label{TCS_225}
\end{eqnarray}
\end{widetext}
Finally, for the sub-process (\ref{H2WW}) we have
\begin{eqnarray}
d\sigma(AB \to l^{+} \nu_l  l^{-} \bar{\nu}_l) &=& exp \left(  {2\pi\over 3}  \alpha_S(\mu_c^2) \right) {dk_{1,t}^2 \over k_{1,t}^2} {dk_{2,t}^2 \over k_{2,t}^2} \; dp_{1,t}^2 \; dp_{2,t}^2 \; dp_{3,t}^2 
 \; dy_1 \; dy_2 \; dy_3 \; dy_4 \; 
 \nonumber \\ & \times & 
 {d\psi_1 \over 2\pi} {d\psi_2 \over 2\pi} {d\phi_1 \over 2\pi} {d\phi_2 \over 2\pi} {d\phi_3 \over 2\pi}
 f_{g}(x_1,k_{1,t}^2, \mu^2) f_{g}(x_2,k_{2,t}^2, \mu^2) 
\nonumber \\ & \times & 
{|\mathcal{M}(gg \to H \to W^+W^- \to l^{+} \nu_l  l^{-} \bar{\nu}_l)|^2 \over 2024\pi^4 (x_1 x_2 s)^2},
\label{TCS_22H24}
\end{eqnarray}
where $\psi_i$ are the azimuthal angles of the initial partons and $\mu_c^2 = m_H^{2/3} p_{H,t}^{4/3}$ with $m_H$ and $p_{H,t}$ being the mass and the transverse momentum of the exchanged Higgs boson in the sub-process (\ref{H2WW}). The functions $f_{a}(x,k_{t}^2,\mu^2)$ in the Eqs. (\ref{TCS_224}), (\ref{TCS_225}) and (\ref{TCS_22H24}) are the KMR UPDFs.  These double-scaled TMD PDFs depend on the fraction of the longitudinal momentum of the parent hadron carried by the incoming parton, $x$, transverse momenta $k_{t}$ and the hard-scale $\mu$. 

The KMR UPDFs are defined as \cite{Kimber:2001sc}
\begin{eqnarray}
f_a(x,k_t^2,\mu^2) = \Delta_S^a(k_t^2,\mu^2) 
\sum_{b=q,g} \left[ {\alpha_S\over 2\pi}
\int^{\mu \over (\mu+k_t)}_{x} dz P_{ab}(z) b\left( {x \over z}, k_t^2 \right) \right] , 
\label{KMR_UPDF}
\end{eqnarray}
with the Sudakov form factor, 
\begin{eqnarray}
\Delta_S^a(k_t^2,\mu^2) =
  exp \left( - \int_{k_t^2}^{\mu^2} {\alpha_S \over 2\pi}
    {dk^{2} \over k^2} \sum_{b=q,g} \int^{\mu \over(\mu+k_t)}_{0} dz' P_{ab}(z') \right),
   \label{SFF}
\end{eqnarray}
with $\alpha_S \equiv \alpha_S(k^2)$ as the running coupling of the strong interaction and $P_{ab}(z)$ as the LO splitting functions for $b \to a+X$ partonic splittings \cite{Kimber:2001sc,Jung:2003wu}. The upper bounds of the integrations in the Eqs. (\ref{KMR_UPDF}) and (\ref{SFF}), i.e. $\mu /(\mu+k_t)$, are the manifestations of the AOC that characterize the kinematics of the KMR UPDFs. Furthermore, $b( x, k_t^2)$ are the solutions of the DGLAP evolution equations. For the purpose of our calculations, these PDFs are obtained from the MMHT2014-LO libraries~\cite{Harland-Lang:2014zoa}. 

Here, we numerically calculate the differential cross-section for the production of $W^+ W^-$ pairs by choosing the hard-scale of the process as
\begin{equation}
    \mu^2 = \sum_{i \in \text{final-state}} p_{i,t}^2,
    \label{HS}
\end{equation}
within the $[0 ,k_{t}^{\text{max}}]$ boundaries for the integrations over the $k_{i,t}$, where
\begin{equation}
    k_{t}^{\text{max}} \equiv 4 \left[ \sum_{i \in \text{final-state}} p_{i,t}^{2,\text{max}} \right]^{1/2},
    \label{ktmax}
\end{equation}
due to the limit ~$f_{a}(x,k_{t}^2\gg\mu^2) \to 0$. Also, for the non-perturbative domain of $k_{i,t} \in [0 , 1 \text{ GeV}]$, $f_{a}(x,k_{t}^2,\mu^2)$ takes on the form 
\begin{equation}
    f_a(x,k_{t}^2<\mu_0^2,\mu^2) = {k_{t}^2 \over \mu_0^2} a(x,\mu_0^2) \Delta_S^a(\mu_0^2,\mu^2), 
\end{equation}
and therefore $\lim_{k_{t}^2 \rightarrow 0} f_{a}(x,k_{t}^2,\mu^2) \sim k_{t}^2$.

The corresponding event selection constraints are chosen by the specifications of the existing experimental measurements as shown in Table~\ref{tab1}.
\begin{table}[t]
\centering
\begin{tabular}{c | c c c c c c c c c}
$\sqrt{s}$ & $p_t^{\text{leading}}$ & \hspace{0.2in} & $p_t^{\text{sub-leading}}$ & \hspace{0.2in} & 
$y_{e \mu}$ & \hspace{0.2in} & $m_{e \mu}$ & \hspace{0.2in} & $E_t^{\text{missing}}$ \\ \hline
$8$~TeV & $>20$~GeV & & $>10$~GeV & & $<\left| 2.5 \right|$ & & $> 10$~GeV & & $> 10$ GeV \\
$13$~TeV & $>27$~GeV & & $>15$~GeV & & $<\left| 2.5 \right|$ & & $> 55$~GeV & & $> 20$~GeV
\end{tabular}
\caption{ \it Event selection criteria for 8 and 13~TeV calculations, compatible with the conditions used in the corresponding ATLAS and CMS measurements. $p_t^{\text{leading}}$ and $p_t^{\text{sub-leading}}$ are the transverse momenta of the leading and the sub-leading leptons, $y_{e \mu}$ and $m_{e \mu}$ are the pseudorapidity and the mass of the dilepton system and $E_t^{\text{missing}}$ is the missing transverse energy due to neutrino emission. In the case of electron production, the pseudorapidity domain is considered to be $\left| y_e \right| < 2.5$ excluding $1.37<\left| y_e \right| < 1.52$.}
\label{tab1}
\end{table}

\section{Numerical results and discussion}
\label{sec:Results}

\begin{figure}[!h]
\centering
\includegraphics[width=1\textwidth]{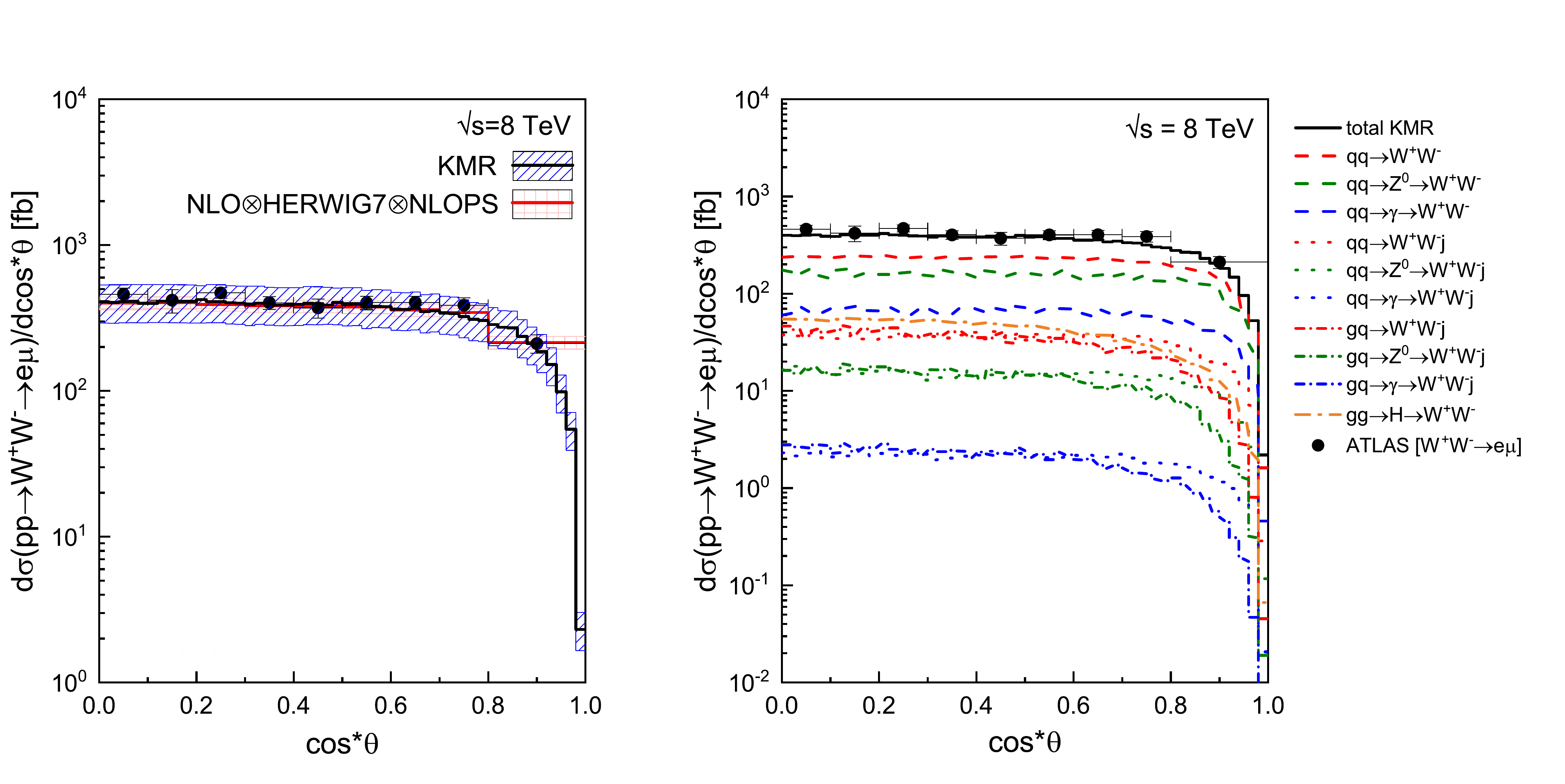}
\caption{ \it Differential cross-section for the production of $W^+W^-$ pairs as a function of $\cos^*\theta$ observable. The calculations were carried out, using the KMR UPDFs at $\sqrt{s}=8$~TeV center-of-mass energy. The right panel demonstrates the contributions of the individual productions channels with the solid black curve showing the overall prediction. The left panel shows a comparison between the results, within the corresponding uncertainty bounds, to the similar predictions from \textsf{Herwig 7} \cite{Bellm:2015jjp} and to the experimental measurements of the ATLAS collaboration \cite{Aad:2016wpd}. To produce the uncertainty region, we have manipulated the hard-scale $\mu$ by a factor of 2.}
\label{cosT-8}
\end{figure}

\begin{figure}[!h]
\centering
\includegraphics[width=1.05\textwidth]{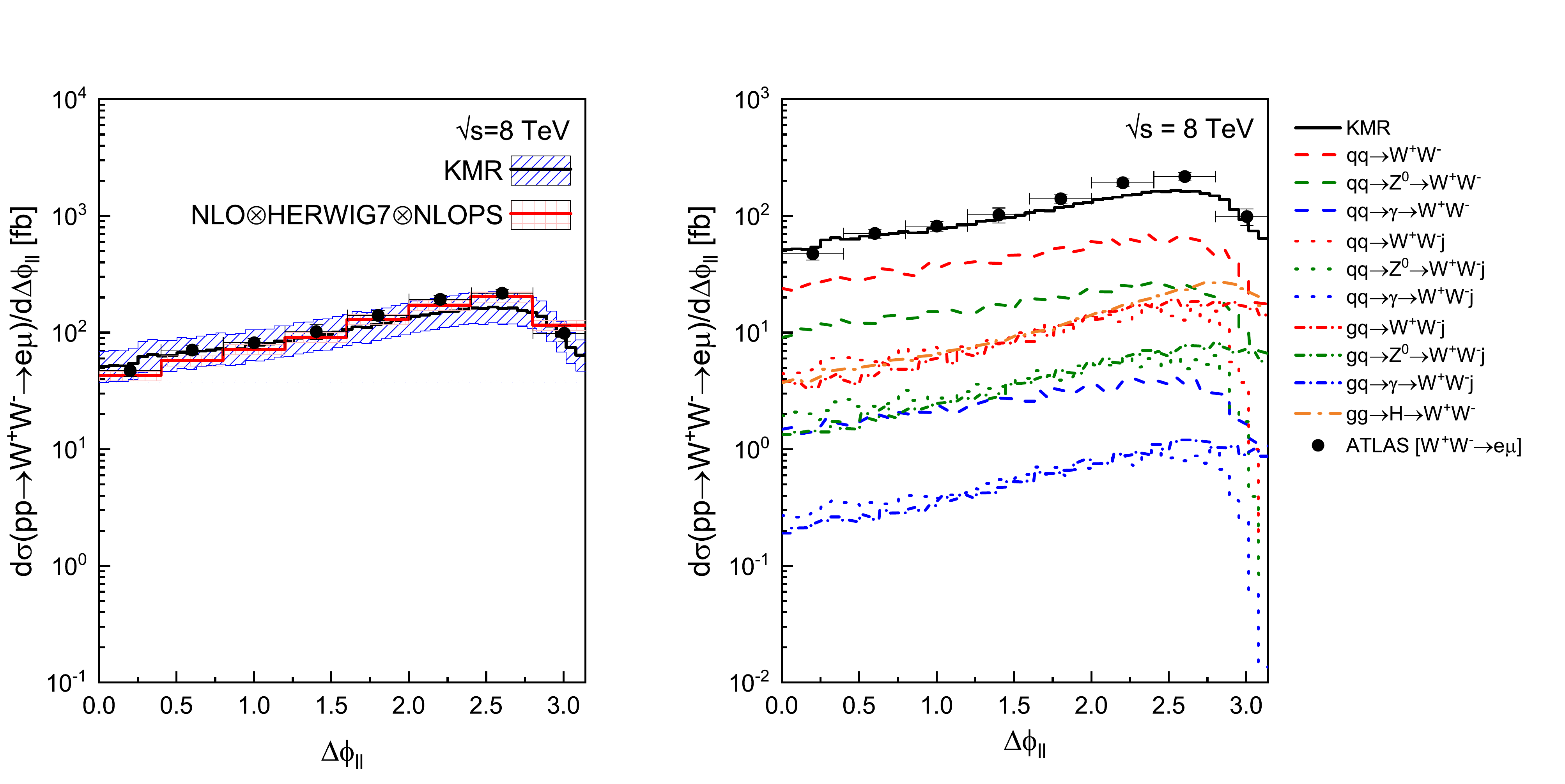}
\caption{ \it Differential cross-section for the production of $W^+W^-$ pairs as a function of the azimuthal angle between the produced leptons, $\Delta \phi_{\ell \ell}$, at $\sqrt{s}=8$~TeV. The notation of the figure is the same as Figure~\ref{cosT-8}.}
\label{DR-8}
\end{figure}

\begin{figure}[!h]
\centering
\includegraphics[width=1.05\textwidth]{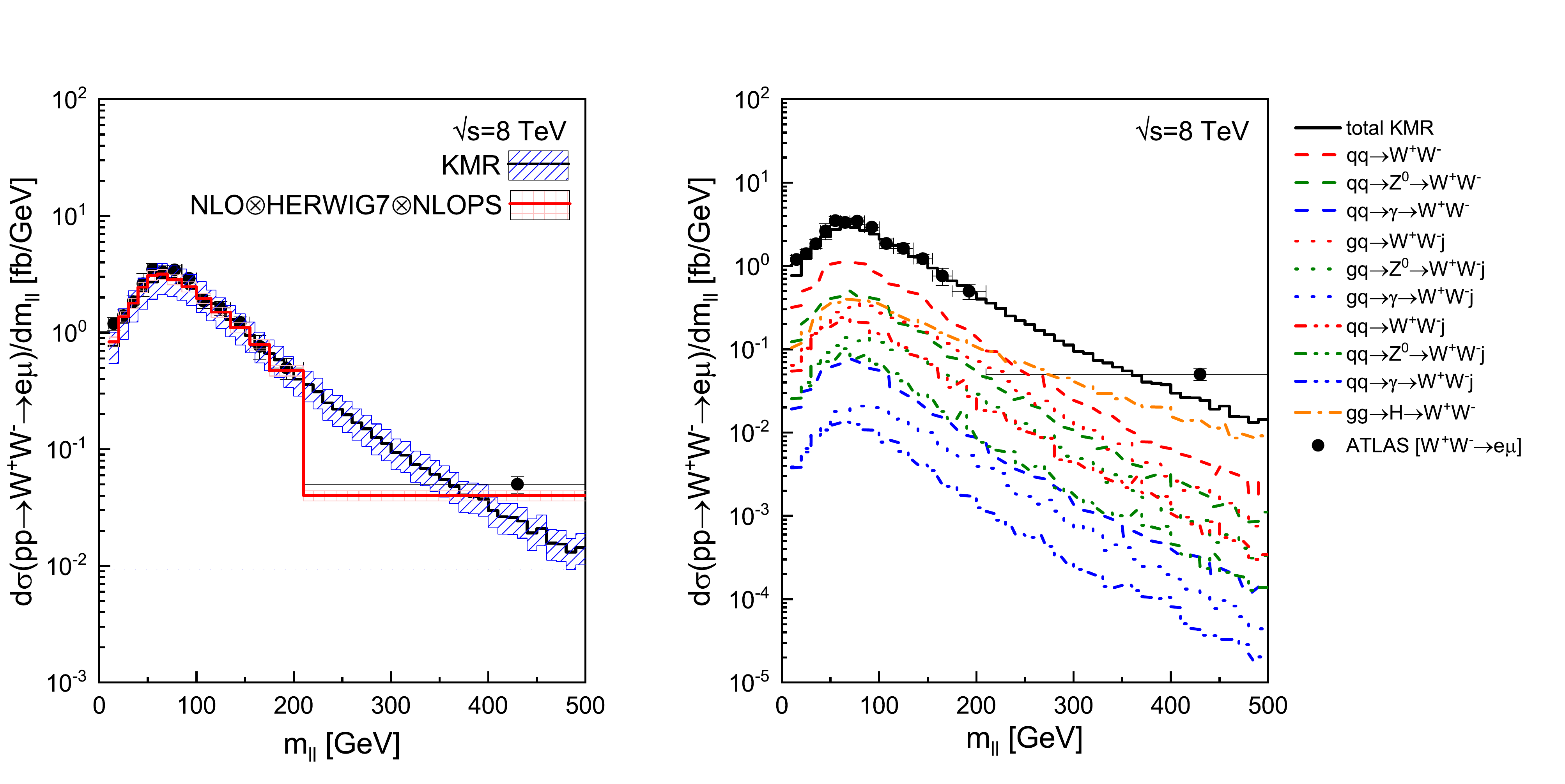}
\caption{ \it Differential cross-section for the production of $W^+W^-$ pairs as a function of the mass of the produced lepton pair, $m_{\ell \ell}$, at $\sqrt{s}=8$~TeV. The notation of the figure is the same as Figure~\ref{cosT-8}.}
\label{mll-8}
\end{figure}

\begin{figure}[!h]
\centering
\includegraphics[width=1.05\textwidth]{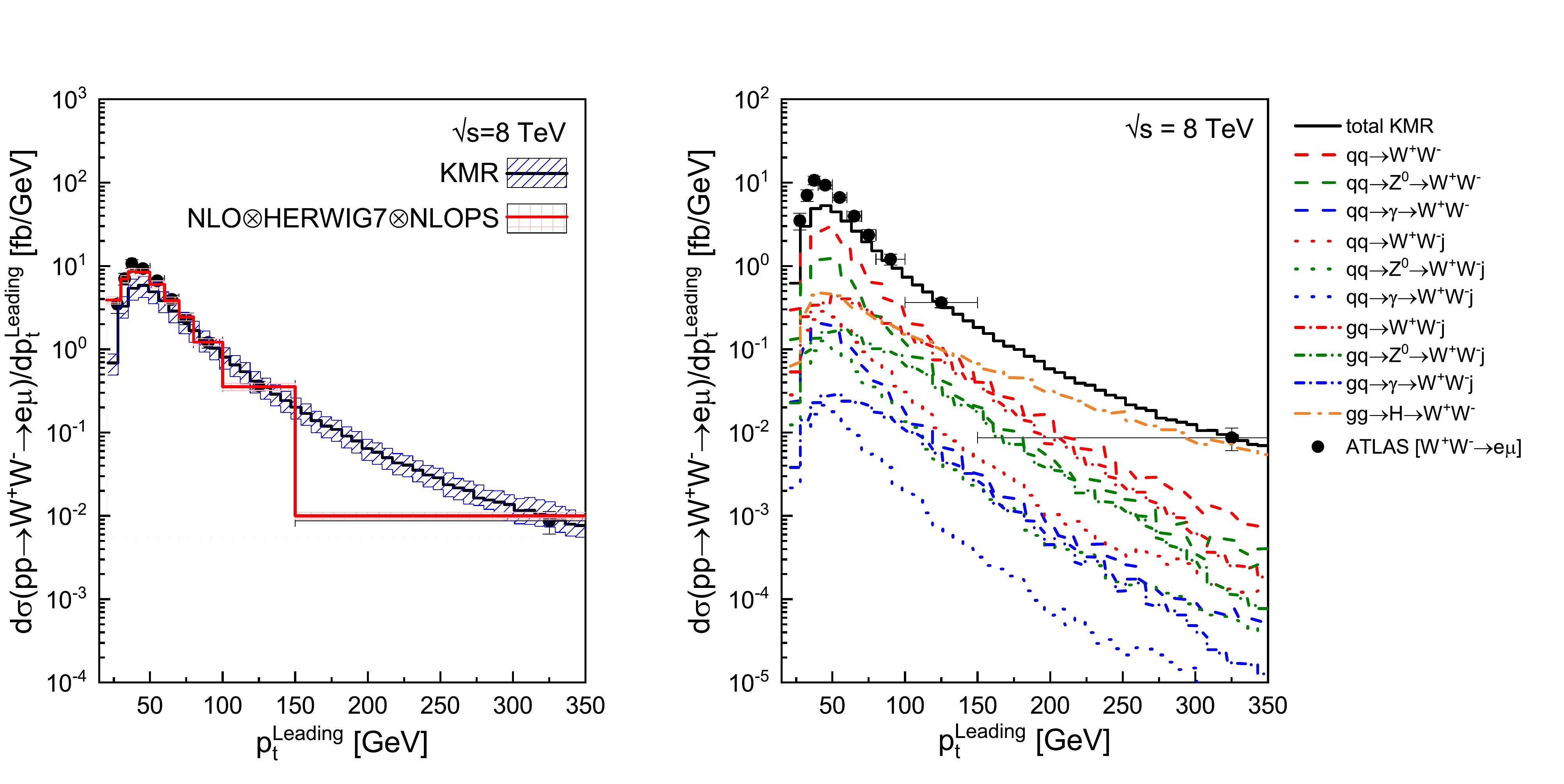}
\caption{ \it Differential cross-section for the production of $W^+W^-$ pairs as a function of the mass of the transverse momentum of the leading lepton, $p_{t}^{\text{Leading}}$, at $\sqrt{s}=8$~TeV. The notation of the figure is the same as Figure~\ref{cosT-8}.}
\label{ptleading-8}
\end{figure}

\begin{figure}[!h]
\centering
\includegraphics[width=1.05\textwidth]{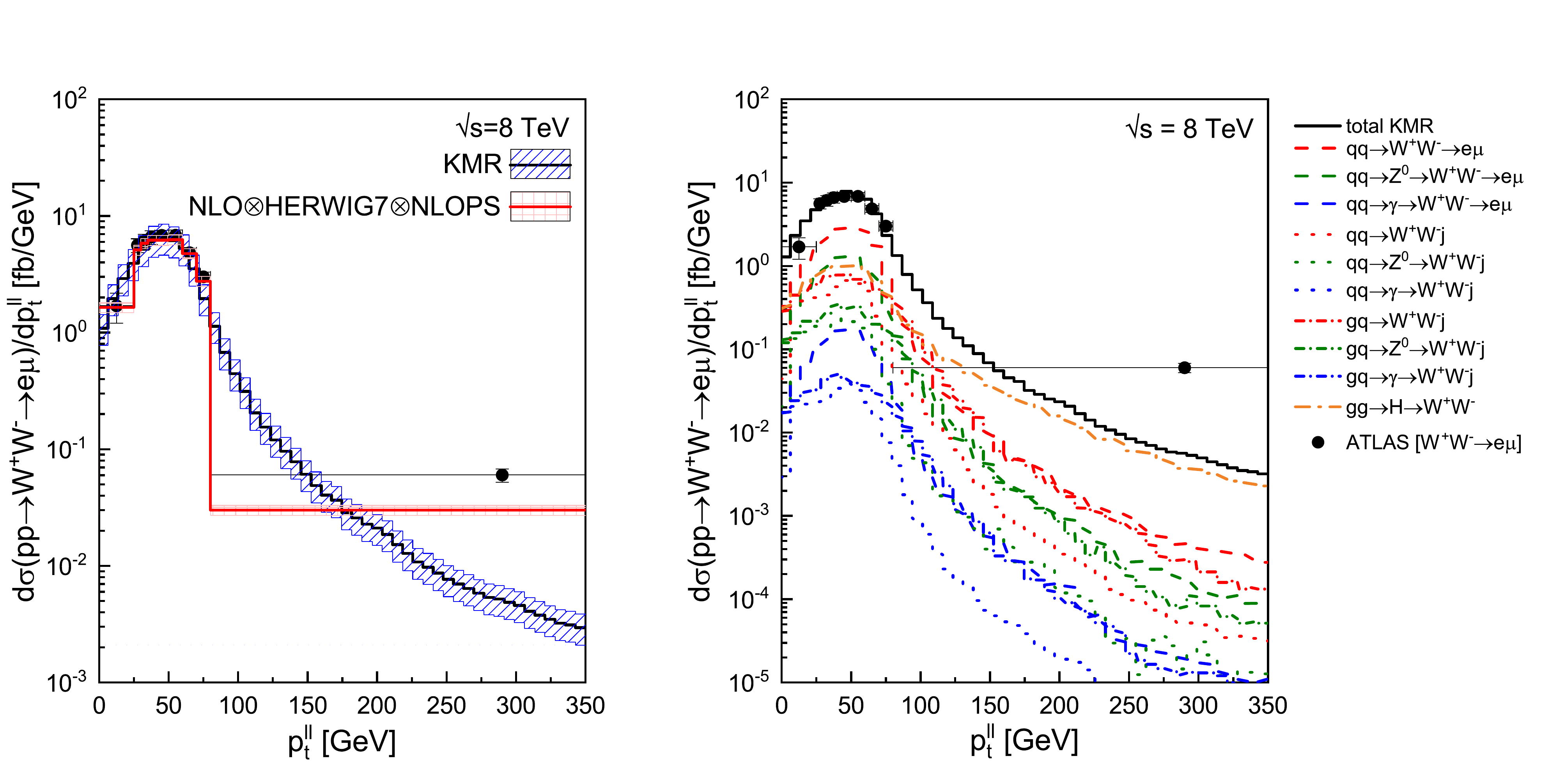}
\caption{ \it Differential cross-section for the production of $W^+W^-$ pairs as a function of the transverse momentum of the produced lepton pair, $p_t^{\ell \ell}$, at $\sqrt{s}=8$~TeV. The notation of the figure is the same as Figure~\ref{cosT-8}.}
\label{pll-8}
\end{figure}

\begin{figure}[!h]
\centering
\includegraphics[width=1.05\textwidth]{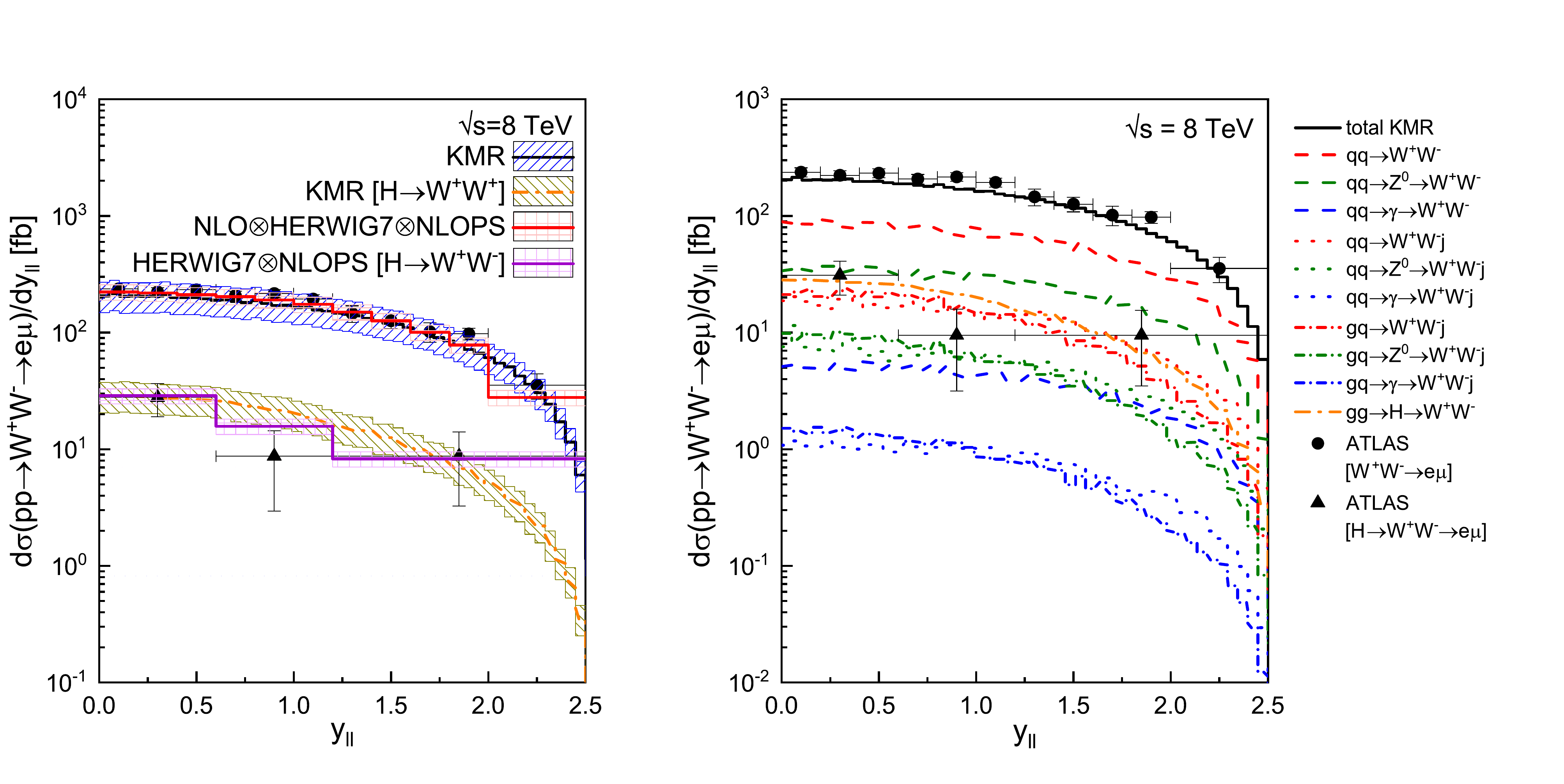}
\caption{ \it Differential cross-section for the production of $W^+W^-$ pairs as a function of the pseudorapidity of the produced lepton pair at $\sqrt{s}=8$~TeV. The right panel demonstrates the contributions of the individual productions channels with the solid black curve showing the overall prediction. The left panel shows a comparison between the results, within the corresponding uncertainty bounds, to the similar predictions from \textsf{Herwig 7} \cite{Bahr:2008pv,Bellm:2015jjp} and to the experimental measurements of the ATLAS collaboration \cite{Aad:2016wpd}. The left panel also contains a similar comparison for $W^+W^-$ pair production via Higgs boson decay and ATLAS \cite{Aad:2016lvc}.}
\label{yll-8}
\end{figure}

In this section, we present our results for the calculation of inclusive $W^{+} W^{-}$ pair production at~$\sqrt{s}=8$~TeV and $13$~TeV, through leptonic decay channels $W^+W^- \to l^+\nu_l + l^{\prime -} \nu_{l'}$ in two different approaches; (i) in the $k_t$-factorization framework using the KMR UPDFs and (ii) in the collinear factorization framework, using the \textsf{Herwig 7} event generator. Our predictions for $W^{+} W^{-}$ pair production at $\sqrt{s}=8$~TeV are shown in Figures~\ref{cosT-8}, \ref{DR-8}, \ref{mll-8}, \ref{ptleading-8}, \ref{pll-8}~and~\ref{yll-8}. The right panels of each of these figures illustrate the individual contributions of the involving sub-processes, with the solid-black histograms representing the total values. The LO contributions are shown by dashed lines, while the LO plus 1 jet histograms are depicted by dotted (short-dashed-dotted) lines corresponding to $q+q \to W^+ + W^- + J$ ($q+g \to W^+ + W^- + J$) channels. The one-loop induced $g+g \to H \to W^+ + W^-$ histograms are plotted as dashed-dotted histograms. In these figures, all tree-level t-channel contributions are coloured in red while the histograms related to the Drell-Yan diagrams associated with an exchanged $Z^0$ ($\gamma$) gauge vector boson are coloured in green (blue). 
Furthermore, the left panels of these figures include the total contributions (labelled as \texttt{KMR}) within the corresponding uncertainty bounds, compared against the existing experimental measurements from the ATLAS and/or CMS collaborations and the NLO predictions in the collinear factorization framework from \textsf{Herwig 7}, labelled with \texttt{NLO$\otimes$HERWIG7$\otimes$NLOPS}, to mark the use of QCD NLO matrix elements and the enhancement via an angularly ordered and MC@NLO matched parton shower. All uncertainty bounds are determined by manipulating the hard-scale $\mu$ by a factor of 2. 

In the figure~\ref{cosT-8} the differential cross-section for the production of $W^+W^-$ pairs are plotted as a function of $\cos^*\theta$,
$$
\left| \cos^*\theta \right| = 
\left| \tanh \left( {\Delta y_{\ell \ell} \over 2} \right) \right|,
$$
with $\Delta y_{\ell \ell}$ being the difference between the rapidities of the produced leptons. As expected, it can be seen that the LO channels have the largest contributions into the production rate ($\sim$62.0\%) with the t-channel $q+q \to W^+ + W^-$ having the greatest impact ($\sim$42.3\%). On the other hand, the LO plus 1 jet channels contribute upto about 27.6\% towards the total cross-section while the one-loop induced $g+g \to H \to W^+ + W^-$ (enhanced by the use of the K-factor approximation) contributes about 10.4\%. The smallest contributions are coming from the s-channel LO plus 1 jet sub-processes that are associated with a $\gamma \to W^+W^-$ decay vertex. These include a 0.5\% share for the quark-quark channel and a 0.6\% for the quark-gluon channel. 

The left panel of Figure~\ref{cosT-8} illustrates a comparison between the $k_t$-factorization predictions, within the corresponding uncertainty bounds, to similar predictions from \textsf{Herwig 7} and the experimental measurements of the ATLAS collaboration \cite{Aad:2016lvc}. In comparison to the experimental measurements, as well as the collinear predictions, it can be observed that the $k_t$-factorization results within their uncertainty bounds produce a relatively good description of the data. 

Similar results are shown in Figures~\ref{DR-8}, \ref{mll-8}, \ref{ptleading-8}, \ref{pll-8} and \ref{yll-8}, where the differential cross-section for $W^+W^-$ pair production is plotted as functions of the difference in the azimuthal angles of the decayed leptons, $\Delta\phi_t^{\ell \ell}$, the invariant mass of the dilepton system, $m_{\ell \ell}$, the transverse momentum of the leading lepton, $p_{t}^{\text{Leading}}$, the transverse momentum of the dilepton system, $p_{t}^{\ell \ell}$ and the pseudorapidity of the dilepton system, $y_{\ell \ell}$, respectively.

The overall behaviour of our predictions in these plots is similar to our previous description for Figure~\ref{cosT-8}, regarding the individual contributions of the relevant sub-processes and their closeness to the experimental measurements as well as the collinear results. What is significant here is the behaviour of the $g+g \to H \to W^+ W^-$ histograms in the high-$p_t$ regions, i.e. for the regions equivalent to $p_t^{\ell \ell} > 150$~GeV (similarly for the region $m_{\ell \ell} > 250$~GeV). In these domains, the $g+g \to H \to W^+ + W^-$ contributions into the production event become dominant, lifting the high-$p_t$ tails of these histograms and ensuring that the total predictions are comparable to the experimental data. 


\begin{figure}[!h]
\centering
\includegraphics[width=0.45\textwidth]{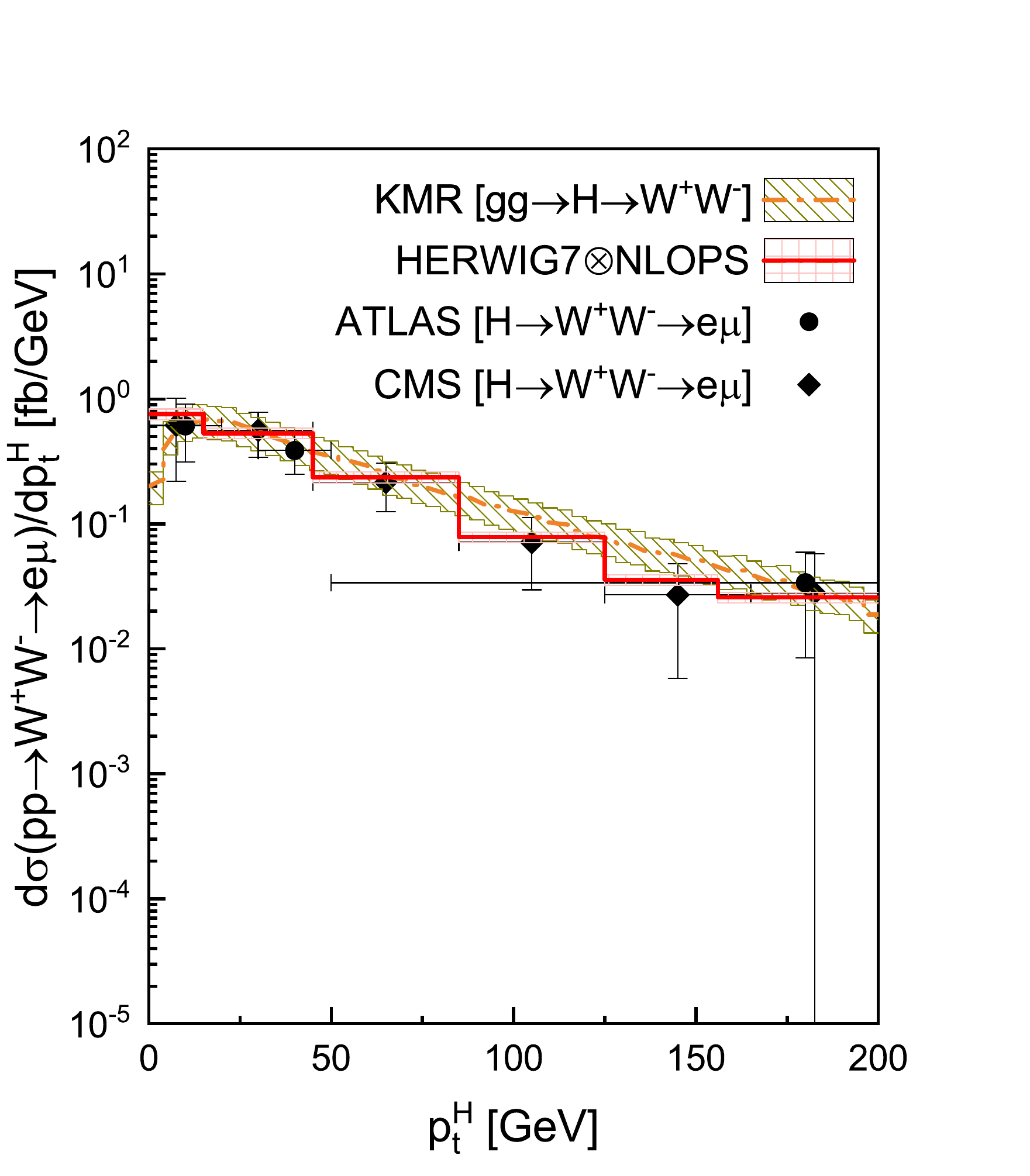}
\caption{ \it Differential cross-section for the production of $W^+W^-$ pairs via Higgs boson decay, as a function of the transverse momentum of the exchanged Higgs boson at $\sqrt{s}=8$~TeV. The plot shows a comparison between the results, within the corresponding uncertainty bounds, to the similar predictions from \textsf{Herwig 7} \cite{Bahr:2008pv,Bellm:2015jjp} and to the experimental data from ATLAS and CMS~\cite{Aad:2016lvc, Khachatryan:2016vnn}.}
\label{ptH-8}
\end{figure}

Besides the above-mentioned indirect deviance for the correct behaviour of our $g+g \to H \to W^+ + W^-$ channel in the the Figures~\ref{DR-8}, \ref{mll-8}, \ref{ptleading-8} and \ref{pll-8}, it is also possible to directly compared our predictions with the experimental measurements for $W^+W^-$ pair production via $H \to W^+ W^-$ decay vertex, from the ATLAS \cite{Aad:2016lvc} and the CMS \cite{Khachatryan:2016vnn} collaborations. Therefore, in addition to inclusive results, the Figure~\ref{yll-8} also compares our one-loop induced $g+g \to H \to W^+ + W^-$ predictions against the experimental data and similar results from \textsf{Herwig 7}. Additionally, in the Figure~\ref{ptH-8}, we have depicted the differential cross-section for the production of $W^+W^-$ pairs via Higgs boson decay, as a function of the transverse momentum of the exchanged Higgs boson. Although in this case, the precision of the data is not as good as the inclusive case, our usual comparisons show a sound behaviour for our predictions within their uncertainty~bounds. 

From the above comparisons, it appears that the SM predictions are satisfactory for describing the experimental image of $W^+W^-$ pair production at the LHC, at least for the $\sqrt{s}=8$~TeV center-of-mass energies. However, one can expect the higher center-of-mass energies to provide a better chance for observing the contributions from BSM, e.g. with extended Higgs sectors. Recently, the ATLAS collaboration has published another set of measurements for the inclusive production of $W^+W^-$ pairs at $\sqrt{s}=13$~TeV \cite{Aaboud:2019nkz}. 

\begin{figure}[!h]
\centering
\includegraphics[width=1.05\textwidth]{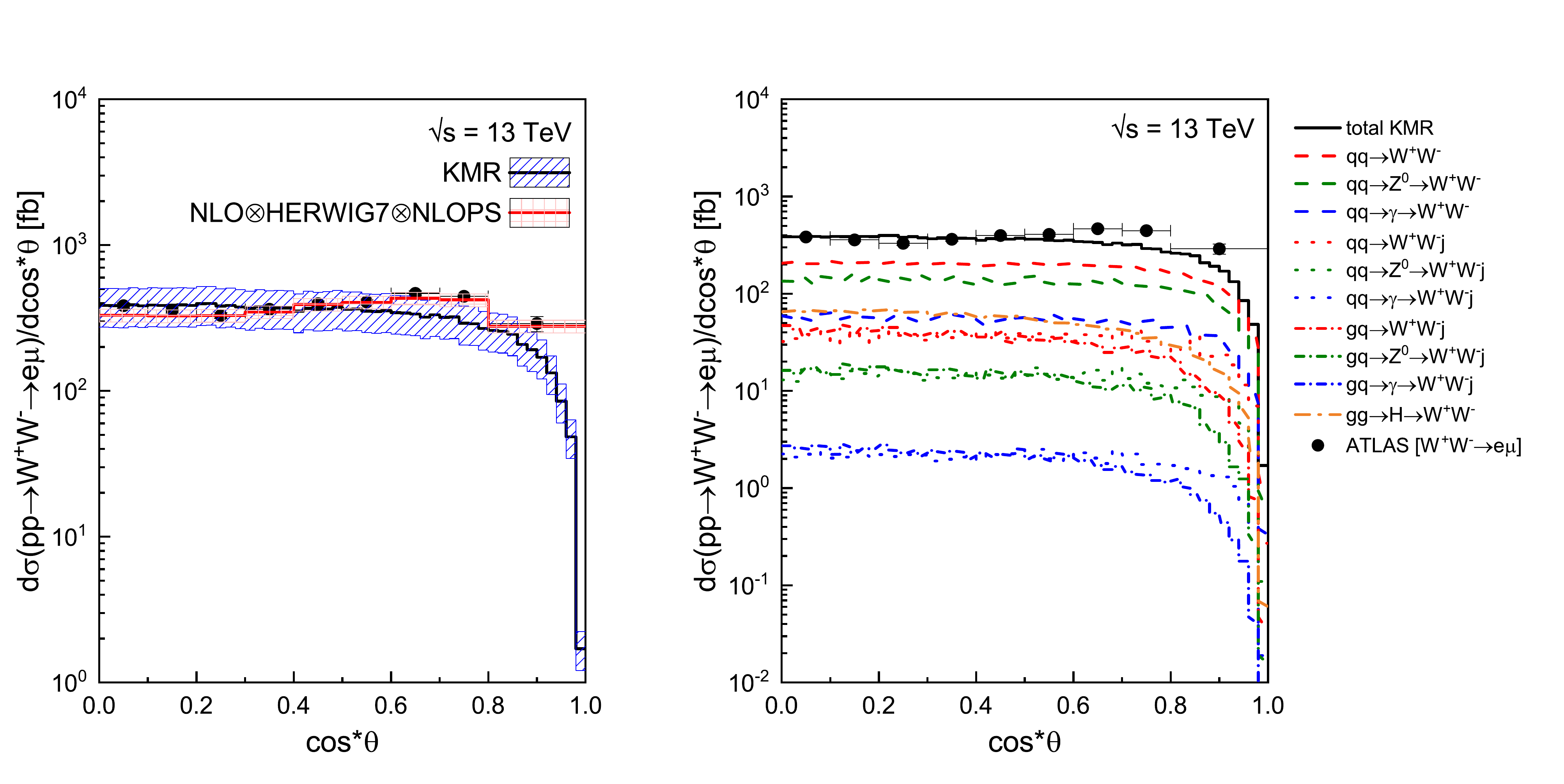}
\caption{ \it Differential cross-section for the production of $W^+W^-$ pairs as a function of the $\cos^*\theta$ observable at $\sqrt{s}=13$~TeV. The right panel demonstrates the contributions of the individual productions channels with the solid black curve showing the overall prediction. The left panel shows a comparison between the results, within the corresponding uncertainty bounds, to the similar predictions from \textsf{Herwig 7} \cite{Bahr:2008pv,Bellm:2015jjp} and to the experimental data from ATLAS \cite{Aaboud:2019nkz}.}
\label{cosT-13}
\end{figure}

\begin{figure}[!h]
\centering
\includegraphics[width=1.05\textwidth]{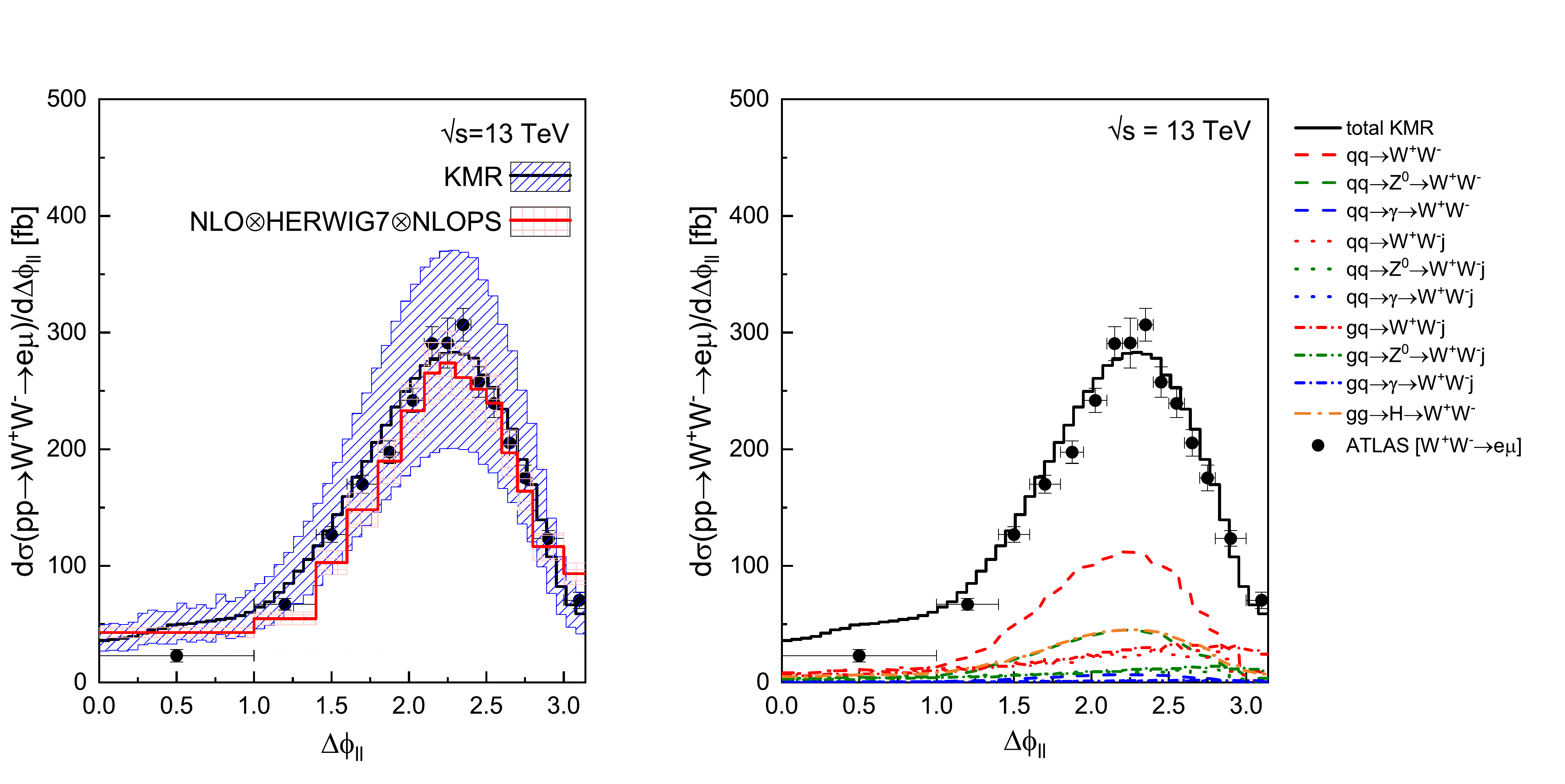}
\caption{ \it Differential cross-section for the production of $W^+W^-$ pairs as a function of the azimuthal angle between the produced leptons, $\Delta \phi_{\ell \ell}$, at $\sqrt{s}=13$~TeV. The notation of the figure is the same as Figure~\ref{cosT-13}.}
\label{DR-13}
\end{figure}

\begin{figure}[!h]
\centering
\includegraphics[width=1.05\textwidth]{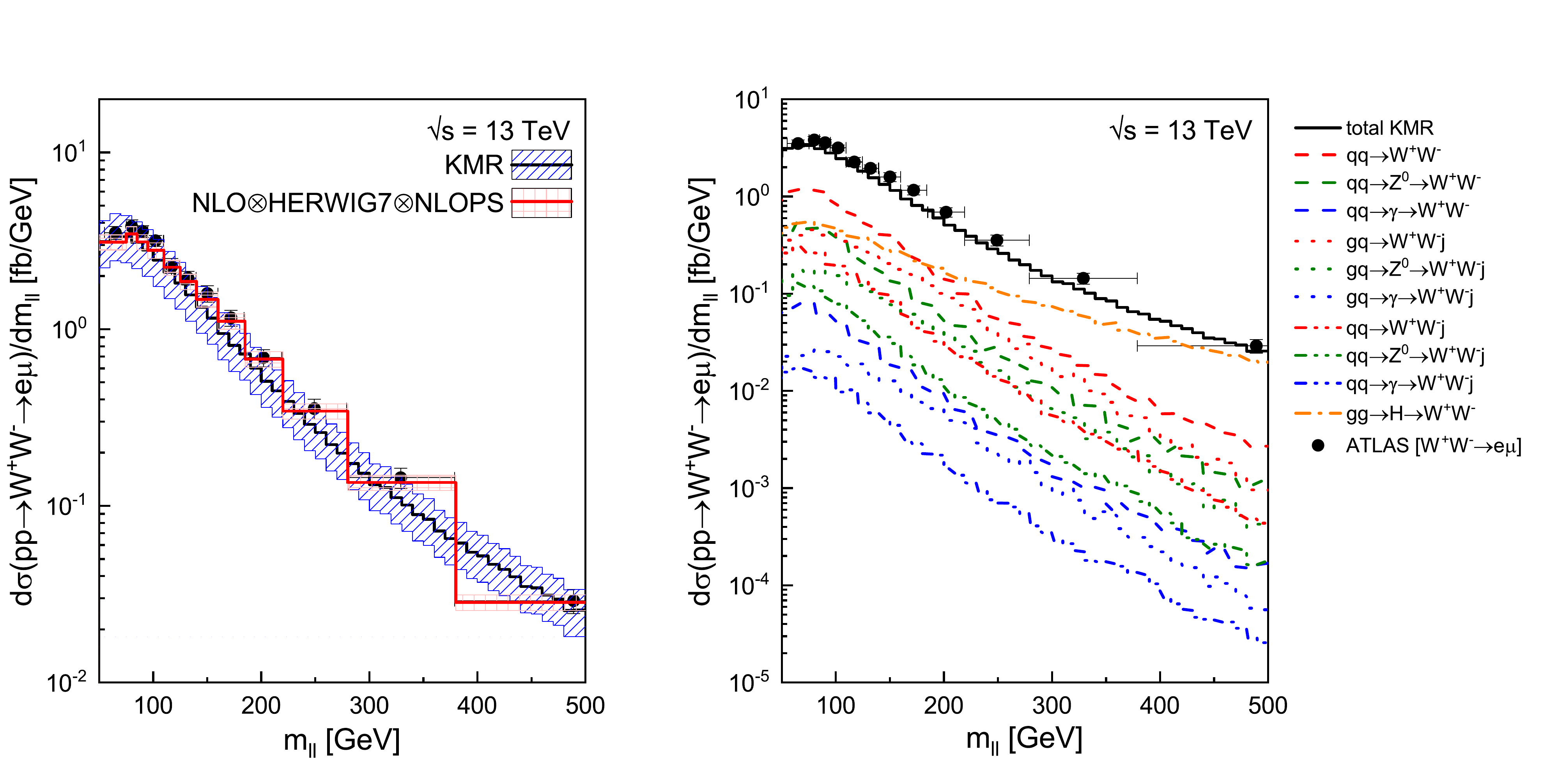}
\caption{ \it Differential cross-section for the production of $W^+W^-$ pairs as a function of the mass of the produced lepton pair, $m_{\ell \ell}$, at $\sqrt{s}=13$~TeV. The notation of the figure is the same as Figure~\ref{cosT-13}.}
\label{mll-13}
\end{figure}

\begin{figure}[!h]
\centering
\includegraphics[width=1.05\textwidth]{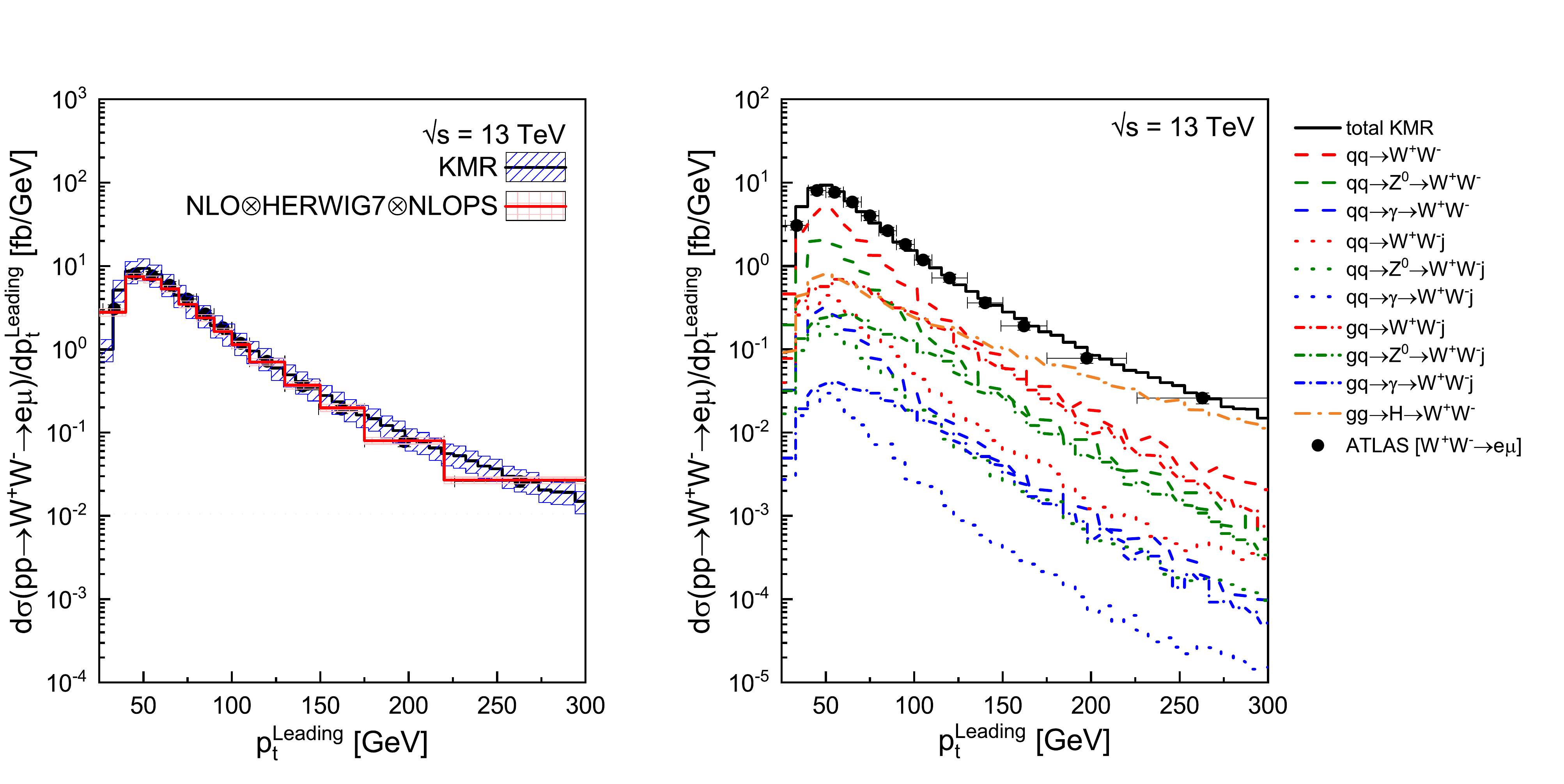}
\caption{ \it Differential cross-section for the production of $W^+W^-$ pairs as a function of the mass of the transverse momentum of the leading lepton, $p_{t}^{\text{Leading}}$, at $\sqrt{s}=13$ TeV. The notation of the figure is the same as Figure~\ref{cosT-13}.}
\label{ptleading-13}
\end{figure}

\begin{figure}[!h]
\centering
\includegraphics[width=1.05\textwidth]{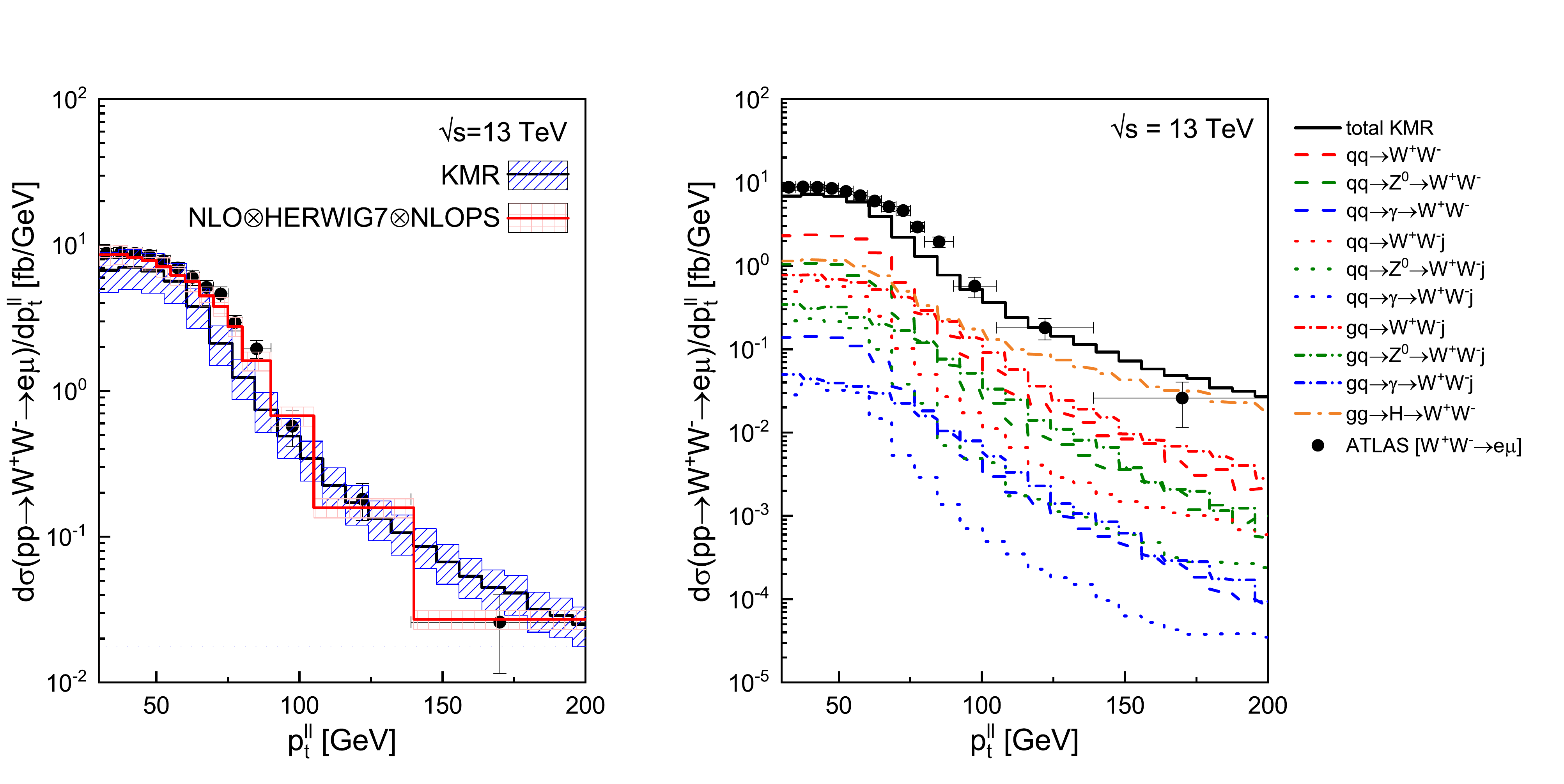}
\caption{ \it Differential cross-section for the production of $W^+W^-$ pairs as a function of the transverse momentum of the produced lepton pair, $p_t^{\ell \ell}$, at $\sqrt{s}=13$~TeV. The notation of the figure is the same as Figure~\ref{cosT-13}.}
\label{pll-13}
\end{figure}

\begin{figure}[!h]
\centering
\includegraphics[width=1.05\textwidth]{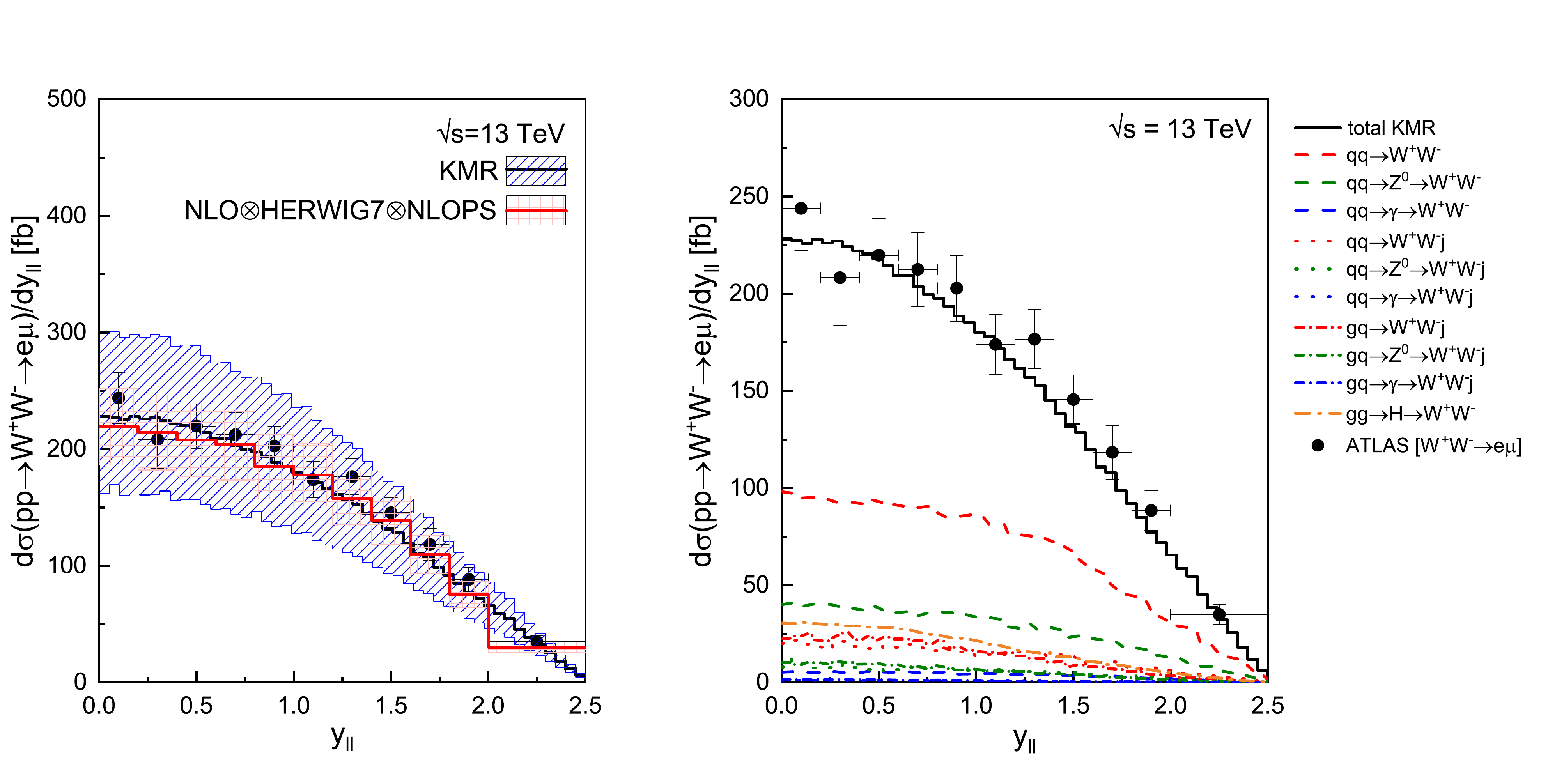}
\caption{ \it Differential cross-section for the production of $W^+W^-$ pairs as a function of the pseudorapidity of the produced lepton pair, $y_{\ell \ell}$, at $\sqrt{s}=13$~TeV. The notation of the figure is the same as Figure~\ref{cosT-13}.}
\label{yll-13}
\end{figure}

In Figures~\ref{cosT-13}, \ref{DR-13}, \ref{mll-13}, \ref{ptleading-13}, \ref{pll-13} and \ref{yll-13}, we have compared our 13~TeV predictions with the similar results from \textsf{Herwig 7} in the collinear factorization framework and with the recent experimental data from \cite{Aaboud:2019nkz}. At first glance, it can be seen that the overall pattern of the behaviour of the sub-processes remains similar to the case of $\sqrt{s}=8$~TeV. A closer investigation, however, reveals that the fractions of the LO plus 1 jet and one-loop induced contributions into the total production rate are slightly greater than the 8~TeV case. At 13~TeV, the LO channels have about a 57.4\% share of the total $W^+W^-$ pair production cross-section, while the LO plus 1 jet $qq$ and $qg$ channels have 13.2\% and 16.7\% shares, respectively. The contribution of the one-loop induced Higgs decay channel is about 12.7\%. 
Note that increasing the center-of-mass energy improves the results of $k_t$-factorization in describing the experimental data. This is since the AOC and an intrinsically embedded initial state real emission\footnote{This is a direct consequence of the last-step evolution approximation and has a similar effect as a single-step initial state parton shower.} are inherently higher-order effects and are more relevant in higher energy contents. 

Overall, the above comparisons demonstrate that our SM base-line calculations in the $k_t$-factorization formalism are satisfactory, in agreement with the conventional theoretical approaches of the same level of QCD accuracy and adequate to describe the experimental~data. 

\section{Conclusions}
\label{sec:Conclusions}
We have calculated the inclusive rate of $W^{+} W^{-}$ pair production through leptonic decay channels $W^+W^- \to l^+\nu_l + l^{\prime -} \nu_{l'}$ in two different approaches; (i) the $k_t$-factorization framework where we used the TMD PDFs of KMR and (ii) the collinear factorization framework, using the \textsf{Herwig 7} event generator. We have performed our analysis with $\sqrt{s}=8$\,TeV and $13$\,TeV center-of-mass energies and compared our predictions with the existing experimental data from ATLAS and CMS.

Here we considered the relevant partonic sub-processes up to a NLO-equivalent-level QCD accuracy. At the $\sqrt{s}=8$~TeV center-of-mass energy, it has been shown that the LO channels have the largest contributions into the production rate ($\sim$62.0\%) while the LO plus 1 jet channels contribute up to about 27.6\% in the total cross-section and the one-loop induced $g+g \to H \to W^+ + W^-$ contributes about 10.4\%. However, for $\sqrt{s}=13$~TeV, the fractions of the LO plus 1 jet and the one-loop induced contributions into the total production rate are slightly greater than the $\sqrt{s}=8$~TeV. In this scale, the LO channels have about a 57.4\% share of the total $W^+W^-$ pair production cross-section, while the LO plus 1 jet $qq$ and $qg$ channels have 13.2\% and 16.7\% shares, respectively. The contribution of the one-loop induced Higgs decay channel is about 12.7\%. 

The $k_t$-factorization framework does not result in high-precision predictions for the $W^+W^-$ pair production event since the UPDFs of $k_t$-factorization has larger uncertainty levels compared to the conventional collinear method, represented here by \textsf{Herwig 7} multi-purpose event generator. \textsf{Herwig 7} provides a versatile platform for performing this class of analysis and incorporates several layers of enhancement to refine its predictions, e.g. NLO QCD corrections, AOC parton shower, multi-particle interactions, hadronization, etc. Nevertheless, we have shown that our $k_t$-factorization framework, despite its simplicity, is adequate for describing the existing experimental data of the fiducial cross-section for the production of $W^+W^-$ pairs in both inclusive and Higgs-decay-originated~cases. 

The large uncertainty of the $k_t$-factorization framework is a consequence of putting the last-step evolution approximation on top of the intrinsic collinear uncertainties and forcing an additional controlling evolution scale ($k_t$) into the calculation. This problem can be solved by performing a global fit for these UPDFs to the Hadron-Hadron scattering data, as well as including higher-order effects in their respective calculations. We hope that the promising results such as those presented in this paper would provide an incentive for the particle physics community to move in this direction.

Additionally, we consider the exclusive $W^{+} W^{-}$ pair production through $gg \to H \to W^+W^-$ channel that is important for the study of the BSM physics. 
Our framework can successfully provide the necessary SM base-line for the on-going search for BSM signal in the LHC run 2 data, which would be the subject of our next paper. 

\begin{acknowledgments}
\noindent
The authors thank Prof. Mike Seymour for reviewing this paper. \textit{MRM} also thanks Prof. Peter Richardson and Dr. Satyajit Seth for their constructive discussions.
\textit{ND} is supported by the Lancaster-Manchester-Sheffield Consortium for Fundamental Physics, under STFC research grant ST/P000800/1. The work of \textit{ND} is also supported in part by HARMONIA grant under contract UMO- 2015/20/M/ST2/00518 (2016-2020).
\textit{MRM} is supported by the UK Science and Technology Facilities Council (grant numbers ST/P001246/1). 
\textit{KO} is supported by the UK Science and Technology Facilities Council (grant numbers ST/N504178/1).
This work has received funding from the European Union's Horizon 2020 research and innovation program as part of the Marie Sk\l{}odowska-Curie Innovative Training Network MCnetITN3 (grant agreement no. 722104).
\end{acknowledgments}


\begin{thebibliography}{a}
\bibitem{Abazov:2009ys}
V.~M.~Abazov {\it et al.} [D0 Collaboration],
Phys.\ Rev.\ Lett.\ {\bf 103} (2009) 191801.
\bibitem{Abazov:2009tr}
  V.~M.~Abazov {\it et al.} [D0 Collaboration],
  Phys.\ Rev.\ D {\bf 80} (2009) 053012.
\bibitem{Aaltonen:2009aa}
 T.~Aaltonen {\it et al.} [CDF Collaboration],
 Phys.\ Rev.\ Lett.\ {\bf 104} (2010) 201801.
\bibitem{Chatrchyan:2011tz}
 S.~Chatrchyan {\it et al.} [CMS Collaboration],
 Phys.\ Lett.\ B {\bf 699} (2011) 25.
\bibitem{Aad:2011kk}
 G.~Aad {\it et al.} [ATLAS Collaboration],
 Phys.\ Rev.\ Lett.\ {\bf 107} (2011) 041802.
\bibitem{Abazov:2011cb}
 V.~M.~Abazov {\it et al.} [D0 Collaboration],
 Phys.\ Rev.\ Lett.\ {\bf 108} (2012) 181803.
\bibitem{Aad:2012rxl}
 G.~Aad {\it et al.} [ATLAS Collaboration],
 Phys.\ Lett.\ B {\bf 712} (2012) 289.
\bibitem{ATLAS:2012mec}
 G.~Aad {\it et al.} [ATLAS Collaboration],
 Phys.\ Rev.\ D {\bf 87} (2013) no.11, 112001
  Erratum: [Phys.\ Rev.\ D {\bf 88} (2013) no.7, 079906].
\bibitem{Chatrchyan:2013oev}
 S.~Chatrchyan {\it et al.} [CMS Collaboration],
 Phys.\ Lett.\ B {\bf 721} (2013) 190.
\bibitem{Chatrchyan:2013yaa}
 S.~Chatrchyan {\it et al.} [CMS Collaboration],
 Eur.\ Phys.\ J.\ C {\bf 73} (2013) no.10, 2610.
\bibitem{ATLAS:2014xea}
 The ATLAS collaboration [ATLAS Collaboration],
 ATLAS-CONF-2014-033.
\bibitem{Khachatryan:2015sga}
 V.~Khachatryan {\it et al.} [CMS Collaboration],
 Eur.\ Phys.\ J.\ C {\bf 76} (2016) no.7,  401.
\bibitem{Aaboud:2017qkn}
  M.~Aaboud {\it et al.} [ATLAS Collaboration],
  Phys.\ Lett.\ B {\bf 773} (2017) 354.
\bibitem{Gallo:2018agq}
  E.~Gallo [CMS Collaboration],
  PoS DIS {\bf 2018} (2018) 083.
\bibitem{Cuevas:2018jah}
  J.~Cuevas [CMS Collaboration],
  PoS LHCP {\bf 2018} (2018) 288.
\bibitem{Aaboud:2018jqu}
  M.~Aaboud {\it et al.} [ATLAS Collaboration],
  Phys.\ Lett.\ B {\bf 789} (2019) 508.
\bibitem{Aad:2016wpd} 
  G.~Aad {\it et al.} [ATLAS Collaboration],
  JHEP {\bf 1609} (2016) 029.
\bibitem{Aad:2016lvc}  
  G.~Aad {\it et al.} [ATLAS Collaboration],
  JHEP {\bf 1608} (2016) 104.
\bibitem{Aaboud:2019nkz}
  M.~Aaboud {\it et al.} [ATLAS Collaboration],
  arXiv:1905.04242 [hep-ex].
\bibitem{Khachatryan:2016vnn}
  V.~Khachatryan {\it et al.} [CMS Collaboration],
  JHEP {\bf 1703} (2017) 032.
\bibitem{Aad:2019lpq}
  G.~Aad {\it et al.} [ATLAS Collaboration],
  arXiv:1903.10052 [hep-ex].
\bibitem{Darvishi:2016fwo}
  N.~Darvishi and M.~R.~Masouminia,
  Nucl.\ Phys.\ B {\bf 923} (2017) 491.
\bibitem{Meade:2014fca}
  P.~Meade, H.~Ramani and M.~Zeng,
  Phys.\ Rev.\ D {\bf 90} (2014) no.11,  114006.
\bibitem{Bai:2012zza}
  Y.~M.~Bai, L.~Guo, X.~Z.~Li, W.~G.~Ma and R.~Y.~Zhang,
  Phys.\ Rev.\ D {\bf 85} (2012) 016008.
\bibitem{Wang:2013qua}
  Y.~Wang, C.~S.~Li, Z.~L.~Liu, D.~Y.~Shao and H.~T.~Li,
  Phys.\ Rev.\ D {\bf 88} (2013) 114017.
\bibitem{Campanario:2013wta}
  F.~Campanario, M.~Rauch and S.~Sapeta,
  Nucl.\ Phys.\ B {\bf 879} (2014) 65.
\bibitem{Bellm:2016cks} 
  J.~Bellm, S.~Gieseke, N.~Greiner, G.~Heinrich, S.~Plätzer, C.~Reuschle and J.~F.~von Soden-Fraunhofen,
  JHEP {\bf 1605}, 106 (2016).
\bibitem{Gehrmann:2014fva}
  T.~Gehrmann, M.~Grazzini, S.~Kallweit, P.~Maierhöfer, A.~von Manteuffel, S.~Pozzorini, D.~Rathlev and L.~Tancredi,
  Phys.\ Rev.\ Lett.\  {\bf 113} (2014) no.21,  212001.
\bibitem{Grazzini:2016ctr}
  M.~Grazzini, S.~Kallweit, S.~Pozzorini, D.~Rathlev and M.~Wiesemann,
  JHEP{\bf~1608}(2016)140.
\bibitem{Kimber:2001sc}
  M.~A.~Kimber, A.~D.~Martin and M.~G.~Ryskin,
  Phys.\ Rev.\ D {\bf 63} (2001) 114027.
\bibitem{Martin:2009ii}
  A.~D.~Martin, M.~G.~Ryskin and G.~Watt,
  Eur.\ Phys.\ J.\ C {\bf 66} (2010) 163.
\bibitem{Gribov:1972ri}
  V.~N.~Gribov and L.~N.~Lipatov,
  Sov.\ J.\ Nucl.\ Phys.\  {\bf 15} (1972) 438
  [Yad.\ Fiz.\  {\bf 15} (1972) 781].
\bibitem{Lipatov:1974qm}
  L.~N.~Lipatov,
  Sov.\ J.\ Nucl.\ Phys.\  {\bf 20} (1975) 94
  [Yad.\ Fiz.\  {\bf 20} (1974) 181].
\bibitem{Altarelli:1977zs} 
  G.~Altarelli and G.~Parisi,
  Nucl.\ Phys.\ B {\bf 126}, 298 (1977).
\bibitem{Dokshitzer:1977sg}
  Y.~L.~Dokshitzer,
  Sov.\ Phys.\ JETP {\bf 46} (1977) 641
   [Zh.\ Eksp.\ Teor.\ Fiz.\  {\bf 73} (1977) 1216].
\bibitem{Deak:2015dpa}
  M.~Deak and K.~Kutak,
  JHEP {\bf 1505} (2015) 068.
\bibitem{Modarres:2018dwj}
  M.~Modarres, M.~R.~Masouminia, R.~Aminzadeh Nik, H.~Hosseinkhani and N.~Olanj,
  Nucl.\ Phys.\ B {\bf 926} (2018) 406.
\bibitem{Modarres:2016tow}
  M.~Modarres, M.~R.~Masouminia, R.~Aminzadeh Nik, H.~Hosseinkhani and N.~Olanj,
  Phys.\ Lett.\ B {\bf 772} (2017) 534.
\bibitem{Modarres:2016phz}
  M.~Modarres, M.~R.~Masouminia, R.~Aminzadeh Nik, H.~Hosseinkhani and N.~Olanj,
  Nucl.\ Phys.\ B {\bf 922} (2017) 94.
\bibitem{AminzadehNik:2018kch}
  R.~Aminzadeh Nik, M.~Modarres and M.~R.~Masouminia,
  Phys.\ Rev.\ D {\bf 97} (2018) no.9,  096012.
\bibitem{Modarres:2016hpe} 
  M.~Modarres, M.~R.~Masouminia, R.~Aminzadeh Nik, H.~Hosseinkhani and N.~Olanj,
  Phys.\ Rev.\ D {\bf 94}, no. 7, 074035 (2016).
\bibitem{Bahr:2008pv}
  M.~Bahr {\it et al.},
  Eur.\ Phys.\ J.\ C {\bf 58} (2008) 639.
\bibitem{Bellm:2015jjp}
  J.~Bellm {\it et al.},
  Eur.\ Phys.\ J.\ C {\bf 76} (2016) no.4,  196.
\bibitem{Bellm:2017bvx} 
  J.~Bellm {\it et al.},
  arXiv:1705.06919 [hep-ph].
\bibitem{Alwall:2014hca}
  J.~Alwall {\it et al.},
  JHEP {\bf 1407} (2014) 079.
\bibitem{Cascioli:2011va}
  F.~Cascioli, P.~Maierhofer and S.~Pozzorini,
  Phys.\ Rev.\ Lett.\  {\bf 108} (2012) 111601.
\bibitem{Buccioni:2017yxi}
  F.~Buccioni, S.~Pozzorini and M.~Zoller,
  Eur.\ Phys.\ J.\ C {\bf 78} (2018) no.1,  70.
\bibitem{Kallweit:2014xda}
  S.~Kallweit, J.~M.~Lindert, P.~Maierhöfer, S.~Pozzorini and M.~Schönherr,
  JHEP{\bf~1504}(2015)012.
\bibitem{Platzer:2011bc}
  S.~Platzer and S.~Gieseke,
  Eur.\ Phys.\ J.\ C {\bf 72} (2012) 2187.
\bibitem{Bellm:2019wrh} 
  J.~Bellm, C.~B.~Duncan, S.~Gieseke, M.~Myska and A.~Siódmok,
  arXiv:1909.08850 [hep-ph].
\bibitem{Bellm:2020}
J. Bellm, D. Grellscheid, P. Kirchgaeßer, A. Papaefstathiou, S. Plätzer, M. Rauch et al., [to appear soon].
\bibitem{Frixione:2002ik}
  S.~Frixione and B.~R.~Webber,
  JHEP {\bf 0206} (2002) 029.
\bibitem{Webber:1983if}
  B.~R.~Webber,
  Nucl.\ Phys.\ B {\bf 238} (1984) 492.
\bibitem{Buckley:2010ar} 
  A.~Buckley, J.~Butterworth, L.~Lonnblad, D.~Grellscheid, H.~Hoeth, J.~Monk, H.~Schulz and F.~Siegert,
  Comput.\ Phys.\ Commun.\  {\bf 184}, 2803 (2013).
\bibitem{Darvishi:2019ltl}
  N.~Darvishi and A.~Pilaftsis,
  Phys.\ Rev.\ D {\bf 99} (2019) no.11,  115014.
\bibitem{Darvishi:2017bhf} 
  N.~Darvishi and M.~Krawczyk,
  Nucl.\ Phys.\ B {\bf 926}, 167 (2018).
\bibitem{Caola:2015rqy}
  F.~Caola, K.~Melnikov, R.~Röntsch and L.~Tancredi,
  Phys.\ Lett.\ B {\bf 754} (2016) 275.
\bibitem{Watt:2003vf} 
  G.~Watt, A.~D.~Martin and M.~G.~Ryskin,
  Phys.\ Rev.\ D {\bf 70}, 014012 (2004)
  Erratum: [Phys.\ Rev.\ D {\bf 70}, 079902 (2004)].
\bibitem{Watt:2003mx}
  G.~Watt, A.~D.~Martin and M.~G.~Ryskin,
  Eur.\ Phys.\ J.\ C {\bf 31} (2003) 73.
\bibitem{Campbell:2006wx}
  J.~M.~Campbell, J.~W.~Huston and W.~J.~Stirling,
  Rept.\ Prog.\ Phys.\  {\bf 70} (2007) 89.
\bibitem{Frixione:1993yp}
  S.~Frixione,
  Nucl.\ Phys.\ B {\bf 410} (1993) 280.
\bibitem{Campbell:2013wga}
  J.~M.~Campbell, R.~K.~Ellis and C.~Williams,
  Phys.\ Rev.\ D {\bf 89} (2014) no.5,  053011.
\bibitem{Baranov:2008hj}
  S.~P.~Baranov, A.~V.~Lipatov and N.~P.~Zotov,
  Phys.\ Rev.\ D {\bf 78} (2008) 014025.
\bibitem{Deak:thesis} M. Deak, 
Ph.D. Thesis, University of Hamburg, Germany, 2009. 
\bibitem{Gribov:1984tu}
  L.~V.~Gribov, E.~M.~Levin and M.~G.~Ryskin,
  Phys.\ Rept.\  {\bf 100} (1983) 1.
\bibitem{Catani:1990eg}
  S.~Catani, M.~Ciafaloni and F.~Hautmann,
  Nucl.\ Phys.\ B {\bf 366} (1991) 135.
\bibitem{Levin:1991ry}
  E.~M.~Levin, M.~G.~Ryskin, Y.~M.~Shabelski and A.~G.~Shuvaev,
  Sov.\ J.\ Nucl.\ Phys.\  {\bf 53} (1991) 657
   [Yad.\ Fiz.\  {\bf 53} (1991) 1059].
\bibitem{Collins:1991ty}
  J.~C.~Collins and R.~K.~Ellis,
  Nucl.\ Phys.\ B {\bf 360} (1991) 3.
\bibitem{FORM} J.A.M. Vermaseren, 
Computer Algebra, 
Nederland, Kruislaan 413, 1098, SJ Amsterdaam, 
1991; ISBN 90-74116-01-9.  
\bibitem{Jung:2003wu}
  H.~Jung,
  Mod.\ Phys.\ Lett.\ A {\bf 19} (2004) 1.
\bibitem{Harland-Lang:2014zoa}
  L.~A.~Harland-Lang, A.~D.~Martin, P.~Motylinski and R.~S.~Thorne,
  Eur.Phys.J.C{\bf~75}(2015)204.
\end{thebibliography}
\end{document}